\documentclass[aip,preprint,byrevtex,nobibnotes,superscriptaddress,floatfix,showkeys]{revtex4-1}

\bibpunct[, ]{[}{]}{;}{a}{,}{,}

\usepackage{graphicx}
\usepackage[caption=false]{subfig}
\usepackage{psfrag}
\usepackage{latexsym}
\usepackage{textcomp}
\usepackage{amssymb}
\usepackage{amsmath}
\usepackage{bm}
\usepackage[Euler]{upgreek}
\usepackage{multirow}
\usepackage{rotating}
\usepackage{hyperref}
\usepackage{epsfig}
\usepackage{dcolumn}
\usepackage{colonequals}
\usepackage[varg]{txfonts}
\usepackage{setspace}

\usepackage{amsmath}
\usepackage{psfrag}
\DeclareMathAlphabet{\mathsf}{OT1}{phv}{b}{n}

\newcommand{\crossVorg}{\ensuremath{%
         \setbox0=\hbox{$V$}
        V \kern-\wd0{\raise.3ex\hbox{$\relbar$}}}}

\newcommand{\crossVxx}[2]{%
	{\setbox0=\hbox{$#1#2V$}
         \setbox1=\hbox{$#1#2$}
         \setbox2=\hbox{$#1V$}
         \dimen1=\wd0
	 \advance\dimen1-\wd1
         \raise.2\ht0\hbox{$#1#2$}\kern-.4\wd0}}

\newcommand{\Figref}[1]{Figure~\ref{#1}}

\newcommand{\degree}{\ensuremath{^\circ}}
\newcommand{\degC}{\ensuremath{\degree\mathrm{C}}}

\usepackage{pifont}

\newcommand{\Rg}{\ensuremath{R_{\mathrm{g}}}}
\newcommand{\Rh}{\ensuremath{R_{\mathrm{H}}}}

\newcommand{\Rgth}{\ensuremath{\Rg^{\theta}}}
\newcommand{\Dth}{\ensuremath{D^{\theta}}}

\newcommand{\urdth}{\ensuremath{U_{\mathrm{RD}}^{\theta}}}

\newcommand{\Rhth}{\ensuremath{R_{\mathrm{H}}^{\theta}}}

\newcommand{\ag}{\ensuremath{\alpha_{\mathrm{g}}}}
\newcommand{\ah}{\ensuremath{\alpha_{\mathrm{H}}}}

\newcommand{\pih}{\ensuremath{\Pi_{\mathrm{H}}}}

\newcommand{\nueff}{\ensuremath{\nu_{\mathrm{eff}}}}

\begin{document}
\title{Universal solvent quality crossover of the zero shear rate viscosity of semidilute DNA solutions }
\date{\today}
\author{Sharadwata Pan}
\affiliation{IITB-Monash Research Academy, Indian Institute of Technology Bombay, Powai, Mumbai - 400076, India}
\affiliation{Department of Chemical Engineering, Indian Institute of Technology Bombay, Powai, Mumbai - 400076, India}
\affiliation{Department of Chemical Engineering, Monash University, Melbourne, VIC 3800, Australia}
\author{Duc At Nguyen}
\affiliation{Department of Chemical Engineering, Monash University,
Melbourne, VIC 3800, Australia}
\author{T. Sridhar}
\affiliation{Department of Chemical Engineering, Monash University, Melbourne, VIC 3800, Australia}
\affiliation{IITB-Monash Research Academy, Indian Institute of Technology Bombay, Powai, Mumbai - 400076, India}
%\author{S. Bhat}
%\affiliation{Polymer Science and Engineering Division, National Chemical Laboratory, Pune, 411 008, India}
\author{P. Sunthar}
\affiliation{Department of Chemical Engineering, Indian Institute of Technology Bombay, Powai, Mumbai - 400076, India}
\affiliation{IITB-Monash Research Academy, Indian Institute of Technology Bombay, Powai, Mumbai - 400076, India}
\author{J. Ravi Prakash}
\email[Corresponding author: ]{ravi.jagadeeshan@monash.edu}
%\homepage[Visit:]{http://users.monash.edu.au/~rprakash/}
\affiliation{Department of Chemical Engineering, Monash University, Melbourne, VIC 3800, Australia}
\affiliation{IITB-Monash Research Academy, Indian Institute of Technology Bombay, Powai, Mumbai - 400076, India}

\begin{abstract}

The scaling behaviour of the zero shear rate viscosity of semidilute unentangled DNA solutions, in the double crossover regime driven by temperature and concentration, is mapped out by systematic experiments. The viscosity is shown to have a power law dependence on the scaled concentration $c/c^{*}$, with an effective exponent that depends on the solvent quality parameter $z$. The determination of the form of this universal crossover scaling function requires the estimation of the $\theta$ temperature of dilute DNA solutions in the presence of excess salt, and the determination of the solvent quality parameter at any given molecular weight and temperature. The  $\theta$ temperature is determined to be $T_\theta \approx 15$\degC\ using static light scattering, and the solvent quality parameter has been determined by dynamic light scattering.

\end{abstract}
\maketitle
%%%%%%%%%%%%%%%%%%%%%%%%%%%%%%%%%%%%%%%%%%%%%%%%%%%%%%
\section{\label{sec:intro}Introduction}
%%%%%%%%%%%%%%%%%%%%%%%%%%%%%%%%%%%%%%%%%%%%%%%%%%%%%%

Many properties of polymer solutions exhibit power law scaling under
$\theta$ solvent and very good solvent conditions. For instance, in
dilute solutions, the radius of gyration \Rg\ scales with molecular
weight $M$ according to the power law $\Rg \sim M^{0.5}$ under
$\theta$ solvent conditions, and $\Rg \sim M^{\nu}$ under very good
solvent conditions, where $\nu \approx 0.59$ is the Flory exponent. In
semidilute solutions, one observes for
instance, ${\eta_{p0}}/{\eta_s} \sim \left({c}/{c^*}\right)^{2}$ in
$\theta$ solvents, while ${\eta_{p0}}/{\eta_s} \sim
\left({c}/{c^*}\right)^{{ 1 }/({3 \nu -1 })}$ in very good
solvents~\citep{dgen79,RubCol03}. Here, $c$ is the polymer mass
concentration, $c^{*}$ is the overlap concentration, which signals the
onset of the semidilute regime, and ${\eta_s}$ and ${\eta_{p0}}$ are
the solvent and zero shear rate polymer contributions to the solution
viscosity, respectively. Power law scaling is, however, \emph{not}
obeyed in the crossover regime between $\theta$ and very good
solvents. Instead, the behaviour of polymer solutions in this regime
is described in terms of universal crossover scaling
functions~\citep{Schafer99}. In the case of dilute polymer solutions,
the nature of these scaling functions is very well understood. Not
only have scaling arguments, analytical theories and computer
simulations established the forms of these scaling functions, they
have been extensively investigated experimentally using a variety of
techniques, and excellent agreement between theory and experiment has
been
demonstrated~\citep{Yama01,MiyFuj81,Schafer99,Hayward19993502,Kumar20037842,SunRav06-epl}. On
the other hand, a comprehensive characterisation of the crossover
scaling functions for semidilute polymer solutions is yet to be
achieved. In this paper, we discuss the systematic measurement of the
crossover scaling function for the zero shear rate viscosity of
semidilute polymer solutions, using DNA molecules as model
polymers. We show that the crossover behaviour of the zero shear rate
viscosity can also be described in terms of a power law, albeit with
an exponent that depends on where the solution lies in the crossover
regime. This behaviour is shown to be in quantitative agreement with
recent Brownian dynamics simulation
predictions~\citep{Jain2012a,Jain2012b}.

The scaling variable that describes the crossover from $\theta$ solvent to very good solvent conditions is the so-called ``solvent quality'' parameter, usually denoted by $z$. While the precise definition of $z$, which depends on both the temperature and the molecular weight, is discussed in greater detail subsequently [see Eq.~(\ref{eq:z})], here it suffices to note that $z=0$ in $\theta$ solvents and $z \to \infty$ in very good solvents, so that the scaling of many dilute polymer solution properties in the crossover regime is typically represented in terms of functions of $z$~\citep{Schafer99,RubCol03}. For instance, the swelling, $\alpha_{\mathrm{g}} = {\Rg (T)} /{\Rgth}$, of the radius of gyration, where, $T$ is the temperature and \Rgth\ is the radius of gyration at the $\theta$ temperature, can be shown to obey the following expression in the crossover regime: $\alpha_{\mathrm{g}} = (1 + a\, z + b\, z^{2} + c \, z^{3})^{m/2}$, where the constants $a$, $b$, $c$, $m$, etc., are either theoretically or experimentally determined constants~\citep{Domb1976179,Schafer99,Kumar20037842}. This expression reduces to the appropriate power laws in the limits $z\to 0$ and $z \to \infty$. The crossover scaling functions for semidilute solutions have an additional dependence on the scaled concentration $c/c^{*}$. We expect, for instance, ${\eta_{p0}}/{\eta_s} = f \left(z, {c}/{c^*}\right)$ in the double crossover regime of temperature and concentration. The specific \emph{power law} forms of these scaling functions in the phase space of solvent quality and concentration, far away from the crossover boundaries, has been predicted previously by scaling theories~\citep{dgen79,GrosbergKhokhlov94,RubCol03}. More recently, using scaling theory based on the blob picture of polymer solutions, Prakash and coworkers~\citep{Jain2012a,Jain2012b} have made a number of predictions regarding the behaviour of scaling functions in the entire $(z, c/c^{*})$ phase space, and, by carrying out Brownian dynamics simulations, have demonstrated the validity of their predictions for the scaling of the polymer size and diffusivity in the semidilute regime. In this work, we investigate experimentally, the scaling of the zero shear rate viscosity of semidilute polymer solutions in the double crossover regime of the variables $z$ and $c/c^{*}$, to examine if the observed scaling behaviour is indeed as predicted by blob scaling arguments. 

Two central conclusions from~\citet{Jain2012a} are of relevance to this work. The first is that there is only one unique scaling function in the double crossover regime of semidilute polymer solutions. In other words, if the scaling function for any one property is known, the scaling function for other properties can be inferred from it. The second conclusion, which comes from the results of Brownian dynamics simulations (since scaling theories cannot predict precise functional forms), is that the crossover scaling functions (in a significant range of values of ${c}/{c^*}$) can also be represented as power laws, but with an effective exponent that depends on $z$. By combining these two observations, one can anticipate that in the semidilute regime, ${\eta_{p0}}/{\eta_s} \sim \left({c}/{c^*}\right)^{{ 1 }/({3 \nueff(z) -1 })}$, where the effective exponent $ \nueff (z)$ is identical to the exponent which characterises the power laws for both the polymer size and the diffusivity. The aim of the experiments carried out here is to establish if such is indeed the case. 
 
In order to examine the scaling behaviour of the zero shear rate viscosity of semidilute polymer solutions in the double crossover regime, it is necessary to measure the viscosity as a function of concentration and temperature for a range of molecular weights, and to represent this behaviour in terms of $z$ and $c/c^{*}$. As is frequently the case in recent studies of polymer solution behaviour, we have used DNA solutions in the presence of excess salt to represent model neutral polymer solutions, because of their excellent monodispersity~\citep{Pecora1991893,smithchu98,Laib20064115,Valle20051,Nayvelt2007477,Ross19681005,Hodnett1976522,Sibileva1987647,Marathias2000153,Nicolai1989,Fujimoto1994304,Robertson2006,Smith1996,Selis1995661,Fishman1996,Chirico1989745,Langowski1987263,Leighton1969313,Doty1958,Liu20091069,Hur2001421,Sunetal05,Schroeder20031515,Babcock20034544}. In spite of the extensive use of DNA solutions, to our knowledge, the $\theta$ temperature of these solutions has not been reported so far.  It is essential to know the $\theta$ temperature in order to describe the temperature crossover of polymer solutions in terms of the scaling variable $z$. In addition, as will be explained in greater detail subsequently, the experimentally determined value of $z$ is arbitrary to within a multiplicative constant. Determining this constant by matching the experimental value of $z$ with the value of $z$ in Brownian dynamics simulations enables a direct comparison of experimentally measured and theoretically predicted crossover scaling functions. 

Static light scattering measurements have been used to determine the $\theta$ temperature of the DNA solutions used in this work. Details of the procedure and the principle results are summarised in Appendix~\ref{sec:sls}. The solvent quality $z$ has been determined by carrying out dynamic light scattering experiments, as described in detail in Appendix~\ref{sec:dls}. Basically, the experimentally measured swelling of the hydrodynamic radius in the temperature crossover regime is mapped onto the results of Brownian dynamics simulations of dilute polymer solutions. The collapse of the data on a master plot demonstrates the universal behaviour of dilute DNA solutions in the presence of excess salt, and enables the determination of $z$ for any combination of temperature $T$ and molecular weight $M$. Section~\ref{sec:method} briefly describes the protocol for our experiments, with details deferred to the supplementary material. The double crossover behaviour of semidilute solutions is examined in section~\ref{sec:zeroshear}. We first demonstrate that at the $\theta$ temperature, the power law scaling ${\eta_{p0}}/{\eta_s} \sim \left({c}/{c^*}\right)^{2}$ is obeyed, as predicted by scaling theory. The dependence of the zero shear rate viscosity on $z$ and $c/c^{*}$ is then examined in the light of the scaling predictions of~\citet{Jain2012a}, and the validity of these predictions in the double crossover regime is established. Finally, we compare measurements of the longest relaxation time $\lambda_{\eta}$ obtained in this work, defined in terms of the zero shear rate viscosity, with the recent measurements of the longest relaxation time $\lambda_{1}$ by Steinberg and coworkers~\citep{Liu20091069}, who observed the relaxation of stained T4 DNA molecules in semidilute solutions following the imposition of a stretching deformation.  The reliability of the current measurements under poor solvent conditions is discussed  in Appendix~\ref{sec:blob}, and our conclusions are summarised in section~\ref{sec:con}.

%%%%%%%%%%%%%%%%%%%%%%%%%%%%%%%%%%%%%%%%%%%%%%%%%%%%%%%%%%%%%%%%%%%%
\section{\label{sec:method}Methodology}
%%%%%%%%%%%%%%%%%%%%%%%%%%%%%%%%%%%%%%%%%%%%%%%%%%%%%%%%%%%%%%%%%%%%

\begingroup
\small
\begin{table*}[t]
%\begin{spacing}{1.2}
  \caption{\label{tab:relaxationtime} Representative properties of DNA
    used in this work. The contour length is estimated using the
    expression $L_{0}$ = number of base-pairs $\times \, 0.34$ nm; the
    molecular weight is calculated from $M$ = number of base-pairs
    $\times \, 662$ g/mol (where the base-pair molecular weight has
    been calculated for a sodium-salt of a typical DNA base-pair segment);
    the number of Kuhn steps from 
    $N_{\mathrm{k}}$ = $L_{0} / (2 P)$ (where $P$ is the persistence
    length, which is taken to be 50 nm), and the radius of gyration at
    the $\theta$ temperature is estimated from $\Rgth\ = L_{0}/\sqrt{6
      N_{\mathrm{k}}}$. The two relaxation times at the $\theta$
    temperature are defined by $\lambda_{\mathrm{D}}^{\theta} =
    {\left(\Rgth\right)}^{2} / \Dth$, where \Dth\ is the measured
    diffusion coefficient under $\theta$ conditions,  and
    $\lambda_{\eta}^{\theta} = (M \eta_{\mathrm{p0}}) / (c
    N_{\mathrm{A}} k_{\mathrm{B}} T)$. While
    $\lambda_{\mathrm{D}}^{\theta}$ is evaluated at $c/c^{*} = 0.1$,
    $\lambda_{\eta}^{\theta}$ is calculated at $c/c^{*} = 1$.} 
\vskip10pt
\begin{tabular}{ c  c  c  c  c  c  c }
\hline
DNA Size (kbp)      & $M$ ($\times 10^{6}$ g/mol)
            & $L_{0} (\mu)$
            & $N_{\mathrm{k}}$
            & $R_{\mathrm{g}}^{\theta}$ (nm)
            & $\lambda_{\mathrm{D}}^{\theta}$ ($\times$10$^{-3}$ s)
            & $\lambda_{\eta}^{\theta}$ ($\times$10$^{-1}$ s)
\\
\hline
\hline
2.96
            & 1.96
            & 1
            & 10
            & 130
            & 7.70
            & --
\\
\hline
5.86
            & 3.88
            & 2
            & 20
            & 182
            & 21.7
            & --
\\
\hline
8.32
            & 5.51
            & 3
            & 28
            & 217
            & 36.9
            & --
\\
\hline
11.1
            & 7.35
            & 4
            & 38
            & 251
            & 56.7
            & --
\\
\hline
25
            & 16.6
            & 9
            & 85
            & 376
            & 197 
            & 1.19
\\
\hline
45
            & 29.8
            & 15
            & 153
            & 505
            & 480
            & --
\\
\hline
48.5
            & 32.1
            & 16
            & 165
            & 524
            & --
            & 4.97
\\
\hline
114.8
            & 76.0
            & 39
            & 390
            & 807
            & 1970
            & --
\\
\hline
165.6
            & 110
            & 56
            & 563
            & 969
            & --
            & 51.9
\\
\hline
289
            & 191
            & 98
            & 983
            & 1280
            & 7930
            & --
\\
\hline
\end{tabular}
\end{table*}
\endgroup

For the purposes of the experiments proposed here, a range of large molecular weight DNA, each with a monodisperse population, is desirable.  This requirement has been met thanks to the work by Smith's group \citep{Laib20064115}, who genetically engineered special double-stranded DNA fragments in the range of 3--300~kbp and incorporated them inside commonly used \emph{Escherichia coli} (\emph{E. coli}) bacterial strains. These strains can be cultured to produce sufficient replicas of its DNA, which can be cut precisely at desired locations to extract the special fragments.

The \emph{E. coli} stab cultures were procured from Smith's laboratory and the DNA fragments were extracted, linearized and purified according to standard molecular biology protocols \citep{Laib20064115,Sambrook2001}.  In addition to the DNA samples procured from Smith's group, two low molecular weight DNA samples (2.9 and 8.3 kbp), procured from Noronha's  laboratory at IIT Bombay, have been used in the light scattering measurements. The various DNA fragments are described in greater detail in the supplementary material. The supplementary material also includes details of working conditions and procedures for preparation and quantification of linear DNA fragments. Table~\ref{tab:relaxationtime} lists some representative properties of all the DNA used here, obtained following the procedures described above.
 
For each molecular weight, the purified linear DNA pellet was dissolved in a solvent (Tris-EDTA Buffer), which is commonly used in experiments involving DNA solutions~\citep{smithetal96, Robertson2006, smithchu98, Sunetal05}. It contains 0.5 M NaCl, which is established (see Appendix~\ref{sec:dls} for details) to be above the threshold for observing charge-screening effects~\citep{marsig95}. Consequently, the DNA molecules are expected to behave identically to neutral molecules. The detailed composition of the solvent is given in the supplementary material.  

The $\theta$-temperature has been determined by carrying out static light scattering measurements with a BI-200SM Goniometer (Brookhaven Instruments Corporation, USA), and the solvent quality parameter $z$ has been determined with the help of dynamic light scattering measurements using a Zetasizer Nano ZS (Malvern, UK), which uses a fixed scattering angle of 173$^{\circ}$. Details of the light scattering measurements, including the sample preparation procedure, and typical scattering intensity plots are given in the supplementary material. 

The viscosity measurements reported here have been carried out on three different DNA molecular weight samples, (i) 25 kbp, procured from Smith's group as described above, (ii) linear genomic DNA of $\lambda$-phage (size 48.5 kbp), purchased from New England Biolabs, U.K. (\#N3011L), and (iii) linear genomic DNA of T4 phage (size 165.6 kbp), purchased from Nippon Gene, Japan (\#314-03973). A Contraves Low Shear 30 rheometer, which is efficient at measuring low viscosities and has very low zero-shear rate viscosity sensitivity at a shear rate of $0.017$ s$^{-1}$~\citep{HeoLarson2005}, was used with cup and bob geometry (1T/1T). The measuring principle of this device has been detailed in an earlier study~\citep{HeoLarson2005}.  One of the primary advantages of using it is the small sample requirement (minimum 0.8 ml), which is ideal for measuring DNA solutions. The zero error was adjusted prior to each measurement. The instrument was calibrated with appropriate Newtonian Standards with known viscosities (around 10, 100 and 1000 mPa-s at 20$^\circ$C) before measuring actual DNA samples. Values obtained fall within 5\% of the company specified values.

Steady state shear viscosities were measured across a temperature range of 10 to 35$^\circ$C for all the linear DNA samples, and a continuous shear ramp was avoided. Prior to measurements, $\lambda$-phage and T4 DNA were kept at 65$^\circ$C for 10 minutes and immediately put in ice for 10 minutes at their maximum concentrations. This was done to prevent aggregation of long DNA chains~\citep{HeoLarson2005}. The shear rate range of the instrument, under the applied geometry, is from 0.01 to 100 s$^{-1}$. At each shear rate, a delay of 30 seconds was employed so that the DNA chains have sufficient time to relax to their equilibrium state. Some typical relaxation times observed in dilute and semidilute solutions are given in Table~\ref{tab:relaxationtime}. At each temperature, a 30 minutes incubation time was employed for sample equilibration.

\begin{table}[t]
\caption{Solvent quality parameter $z$ and overlap concentration $c^{*}$ (in mg/ml) for all the DNA, at various temperatures. The $\theta$-temperature is taken to be 15\degC. } % title name of the table
\vskip10pt
\centering % centering table
\begin{tabular}{l  c c c ccccc} % creating 9 columns 
\hline\hline % inserting double-line 
 &  &  & & 15\degC & 20\degC  & 25\degC & 30\degC & 35\degC \\ [0.5ex]
\hline % inserts single-line
% Entering 1st row 
 &  & $z$ & & 0 & 0.11 & 0.22 & 0.32 &0.43  \\[-1ex]
\raisebox{1.5ex}{2.9 kbp}  &  & $c^{*}$ & & 0.371 & 0.313 & 0.278 & 0.253 & 0.234  \\[1ex]
% Entering 2nd row 
&  & $z$ & & 0 & 0.16 & 0.31 & 0.46 &0.60  \\[-1ex]
\raisebox{1.5ex}{5.9 kbp}  &  & $c^{*}$ &  & 0.251 & 0.201 & 0.173 & 0.155 & 0.142  \\[1ex]
% Entering 3rd row 
&  & $z$ & & 0 & 0.19 & 0.37 & 0.54 &0.71  \\[-1ex]
\raisebox{1.5ex}{8.3 kbp}  &  & $c^{*}$ &  & 0.214 & 0.165 & 0.141 & 0.125 & 0.114  \\[1ex]
% Entering 4th row 
&  & $z$ & & 0 & 0.22 & 0.43 & 0.63 &0.83  \\[-1ex]
\raisebox{1.5ex}{11.1 kbp}  &  & $c^{*}$ &  & 0.184 & 0.139 & 0.117 & 0.103 & 0.093  \\[1ex]
% Entering 5th row 
&  & $z$ & & 0 & 0.33 & 0.64 & 0.95 &1.24  \\[-1ex]
\raisebox{1.5ex}{25 kbp}  &  & $c^{*}$ &  & 0.123 & 0.084 & 0.068 & 0.059 & 0.052  \\[1ex]
% Entering 6th row 
&  & $z$ & & 0 & 0.44 & 0.86 & 1.27 & 1.66  \\[-1ex]
\raisebox{1.5ex}{45 kbp}  &  & $c^{*}$ &  & 0.092 & 0.058 & 0.045 & 0.039 & 0.034  \\[1ex]
% Entering 7th row 
&  & $z$ & & 0 & 0.69 & 1.37 & 2.03 &2.66  \\[-1ex]
\raisebox{1.5ex}{114.8 kbp}  &  & $c^{*}$ &  & 0.057 & 0.031 & 0.023 & 0.019 & 0.017  \\[1ex]
% Entering 8th row 
&  & $z$ & & 0 & 1.11 & 2.18 & 3.22 & 4.22  \\[-1ex]
\raisebox{1.5ex}{289 kbp}  &  & $c^{*}$ &  & 0.036 & 0.016 & 0.012 & 0.010 & 0.008  \\[1ex]
% [1ex] adds vertical space
\hline % inserts single-line 
\end{tabular}
\label{tab:zc}
\end{table}
%\begin{figure}[t]
%\centering
%\begin{tabular}{cc}
%\includegraphics[width=0.555\linewidth]{rawvisc1.eps} & 
%\includegraphics[width=0.515\linewidth]{rawvisc2.eps} \\
%(a) & (b)  \\
%\end{tabular}
%\label{fig:etatemp}
%\caption{}
%\end{figure}
%%
\begin{figure}[!ht]
\centering 
\subfloat[][]{ 
    \includegraphics[width=0.65\textwidth]{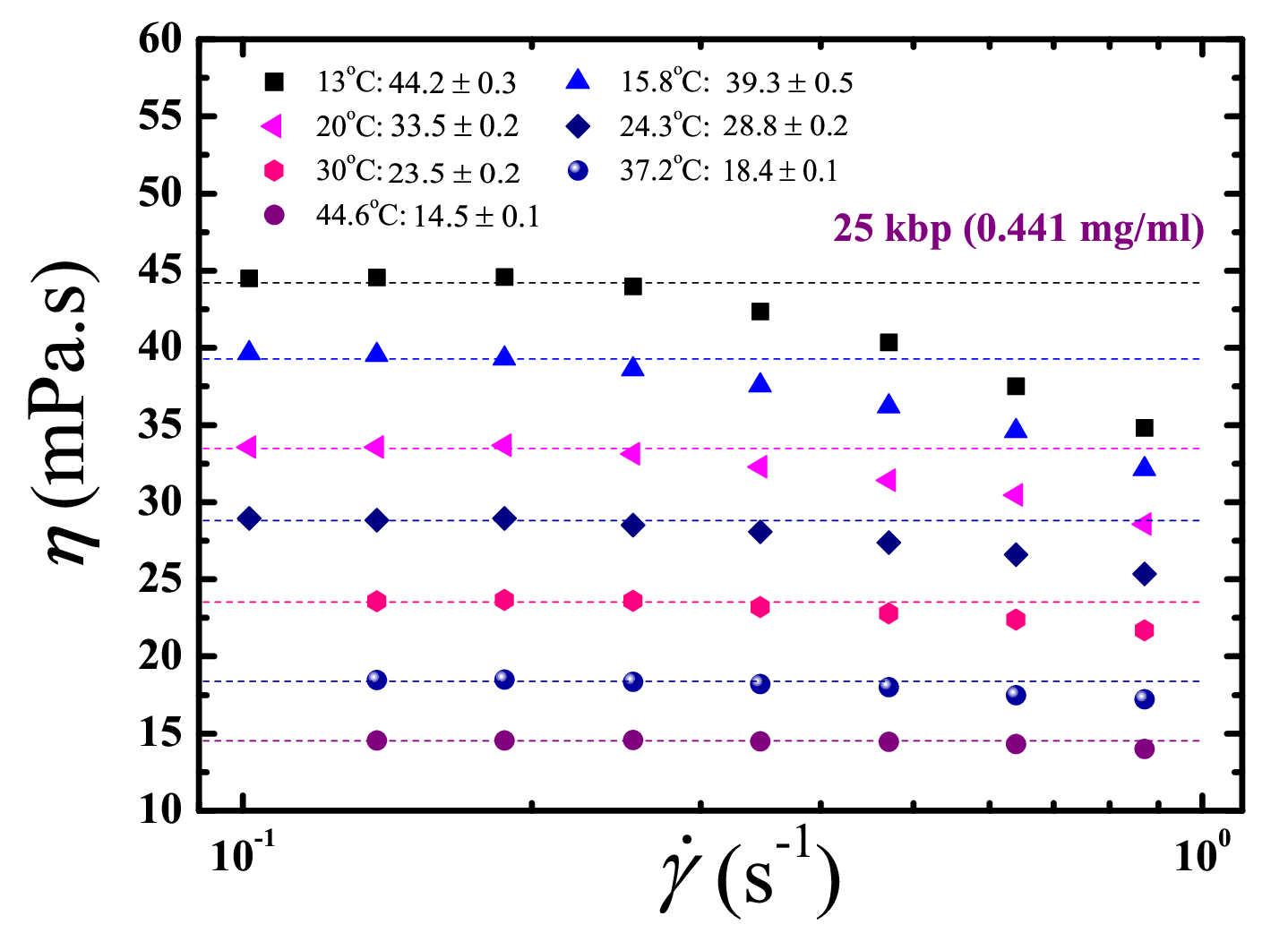}
  \label{figvisc:sub:a}}    \\
\subfloat[][]{ 
    \includegraphics[width=0.63\textwidth]{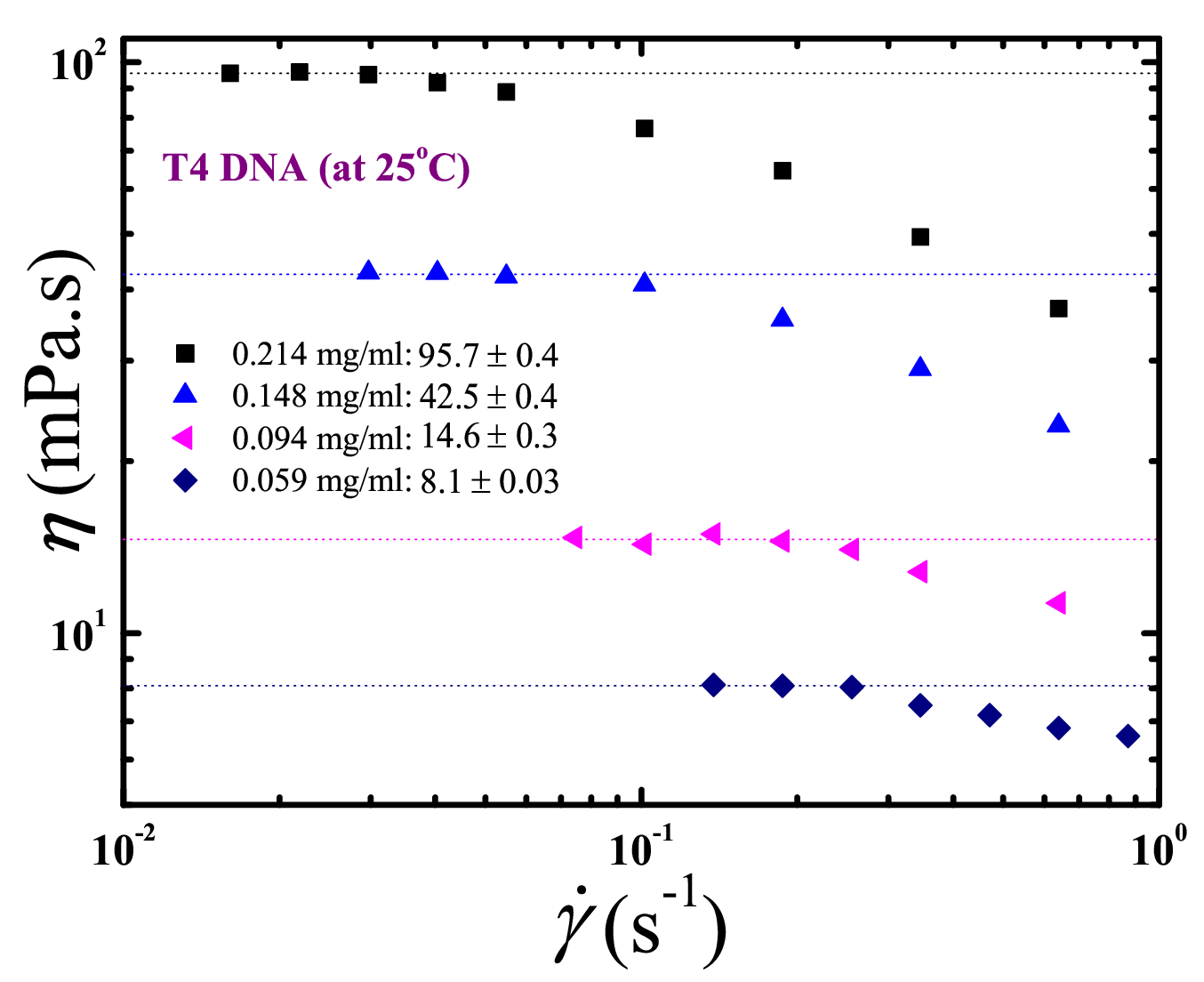}
   \label{figvisc:sub:b}}
%\begin{spacing}{1.5}
\caption{\label{fig:etatemp} Determination of the zero shear rate viscosity. The shear rate dependence of viscosity in the region of low shear rate is extrapolated to zero shear rate: (a) at a fixed concentration, for a range of temperatures, and, (b) at a fixed temperature, for a range of concentrations. Legends indicate the extrapolated values in the limit of zero shear rate.}
%\end{spacing}
\end{figure}

\section{\label{sec:zeroshear}Solvent quality crossover of the zero shear rate viscosity}

\subsection{\label{sec:swelling} Zero shear rate viscosity of semidilute solutions}

The scaling behaviour of the zero shear rate viscosity of semidilute polymer solutions can be determined by measuring the viscosity as a function of concentration and temperature for a range of molecular weights, and then representing this behaviour in terms of the crossover  variables $z$ and $c/c^{*}$. In order to do so, however, as discussed earlier in section~\ref{sec:intro}, it is first necessary to determine the $\theta$-temperature, the solvent quality parameter $z$, and the overlap concentration $c^{*}$. We show in Appendix~\ref{sec:sls}, with the help of static light scattering experiments, that the $\theta$ temperature of the DNA solutions used here is $14.7\pm0.5~\degC$. We have used $T_{\theta}=15 \degC$ in all the calculations carried out here, since we have measurements at this temperature. The solvent quality $z$, which is a function of molecular weight and temperature, is determined with the help of dynamic light scattering experiments, as detailed in Appendix~\ref{sec:dls}. Representative values of $z$, obtained by this procedure at various values of $M$ and $T$, are displayed in Table~\ref{tab:zc}. 

The overlap concentration is defined by the expression $c^{*} = M / \left[(4 \pi/3) \, \Rg^{3} \,N_{A} \right]$, where $N_{A}$ is the Avogadro number. The radius of gyration can be determined from the expression $\Rg = \Rgth \, \ag (z)$, for any $M$ and $T$. Since the chain confirmations at the $\theta$ temperature are expected to be ideal Gaussian chains, the analytical value for the radius of gyration at $T_{\theta}$ is, $\Rgth = L_{0}/\sqrt{6 N_{\mathrm{k}}}$. We have consequently used the respective values of $L_{0}$ and $N_{\mathrm{k}}$ for all the molecular weights used here, to determine \Rgth\ (as displayed in Table~\ref{tab:relaxationtime}). Further, since we know $z$, \ag\ can be determined from the expression $\alpha_{\mathrm{g}} = (1 + a\, z + b\, z^{2} + c \, z^{3})^{m/2}$, where the constants, $a = 9.528$, $b = 19.48$, $c = 14.92$, and $m = 0.1339$ have been determined earlier by Brownian dynamics simulations~\citep{Kumar20037842}. Note that we expect the estimated values of \Rg\ to be close to the actual values for DNA,  since measured crossover values for the hydrodynamic radius \Rh\ agree with the results of Brownian dynamics simulations at identical values of $z$ (as demonstrated in Appendix~\ref{sec:dls}).  Representative values of $c^{*}$ found using this procedure, at various $M$ and $T$, are displayed in Table~\ref{tab:zc}. 

Figures~\ref{fig:etatemp} (a) and (b) display examples of the dependence of the measured steady state shear viscosity on the shear rate. As indicated in the figures, values of viscosity in the plateau region of very low shear rates, at each temperature and concentration, were least-square fitted with a straight line and extrapolated to zero shear rate, in order to determine the zero shear rate viscosities. All the zero shear rate viscosities determined in this manner, across the range of molecular weights, temperatures and concentrations examined here, are displayed in Table~\ref{tab:allvisc}.

%\begingroup
%\squeezetable
\begin{table*}[h]
\caption{\label{tab:allvisc} Zero shear rate steady state viscosities (mPa.s) for 25 kbp, $\lambda$-phage, and T4 DNA at various concentrations (mg/ml) and temperatures (\degC) in the semidilute regime.}
\begin{minipage}[t]{.33\linewidth}
\vspace{0pt}
\centering
\begin{tabular}{ l  l c  c }
\hline
\multicolumn{4}{c}{25 kbp}\\
\hline
$c$       & T           & $c/c^{*}$ & $\eta_{0}$ \\ [0.5ex]
\hline
\hline
\multirow{7}{*}{0.441} & 13 & 2.12 & 44.2 $\pm$ 0.3 \\
      & 15.8 & 3.9 & 39.3 $\pm$ 0.5 \\
      & 20 & 5.13 & 33.5 $\pm$ 0.2 \\
      & 24.3 & 6.04 & 28.8 $\pm$ 0.2 \\
      & 30 & 7 & 23.5 $\pm$ 0.2 \\
      & 37.2 & 7.88 & 18.4 $\pm$ 0.1 \\
      & 44.6 & 8.65 & 14.5 $\pm$ 0.1 \\
\hline
\multirow{7}{*}{0.364} & 13 & 1.75 & 22.6 $\pm$ 0.1 \\
      & 15.8 & 3.22 & 20.4 $\pm$ 0.1 \\
      & 20 & 4.23 & 17.7 $\pm$ 0.1 \\
      & 24.3 & 4.99 & 15.5 $\pm$ 0.1 \\
      & 30 & 5.78 & 12.8 $\pm$ 0.1 \\
      & 37.2 & 6.5 & 10.5 $\pm$ 0.1 \\
      & 44.6 & 7.14 & 8.5 $\pm$ 0.01 \\
\hline
\multirow{7}{*}{0.315} & 13 & 1.51 & 15.8 $\pm$ 0.1 \\
      & 15.8 & 2.79 & 14.7 $\pm$ 0.04 \\
      & 20 & 3.66 & 12.7 $\pm$ 0.1 \\
      & 24.3 & 4.32 & 11.2 $\pm$ 0.01 \\
      & 30 & 5 & 9.3 $\pm$ 0.03 \\
      & 37.2 & 5.63 & 7.7 $\pm$ 0.05 \\
      & 44.6 & 6.18 & 6.6 $\pm$ 0.05 \\
\hline
\multirow{5}{*}{0.112} & 18 & 1.18 & 2.7 $\pm$ 0.02 \\
      & 21 & 1.37 & 2.5 $\pm$ 0.01 \\
      & 25 & 1.56 & 2.3 $\pm$ 0.02 \\
      & 30 & 1.78 & 2 $\pm$ 0.02 \\
      & 35 & 1.93 & 1.8 $\pm$ 0.01 \\
\hline
\multirow{2}{*}{0.07}  & 30 & 1.11 & 1.5 $\pm$ 0.01 \\
      & 35 & 1.21 & 1.5 $\pm$ 0.01 \\
\hline
% \vskip-10pt
\end{tabular}
\end{minipage}
\begin{minipage}[t]{.33\linewidth}
\vspace{0pt}
\centering
\begin{tabular}{ l  l  c  c }
\hline
\multicolumn{4}{c}{$\lambda$ DNA}\\
\hline
$c$       & T           & $c/c^{*}$ & $\eta_{0}$ \\  [0.5ex]
\hline
\hline
\multirow{4}{*}{0.5}   & 10 & -- & 408.7 $\pm$ 9.7 \\
      & 13 & -- & 357.7 $\pm$ 5.9 \\
      & 21 & 9.26 & 334.8 $\pm$ 10 \\
      & 25 & 10.87 & 291.2 $\pm$ 15.1 \\
\hline
\multirow{4}{*}{0.315} & 10 & -- & 82.6 $\pm$ 0.5 \\
      & 13 & -- & 71.5 $\pm$ 2.1 \\
      & 21 & 5.83 & 61.4 $\pm$ 1.05 \\
      & 25 & 6.85 & 57.9 $\pm$ 0.9 \\
\hline
\multirow{7}{*}{0.2}   & 10 & -- & 19.4 $\pm$ 0.2 \\
      & 13 & -- & 16.2 $\pm$ 0.7 \\
      & 15 & 2.25 & 16 $\pm$ 0.1 \\
      & 21 & 3.7 & 14.6 $\pm$ 0.3 \\
      & 25 & 4.35 & 12.3 $\pm$ 0.6 \\
      & 30 & 4.88 & 11.3 $\pm$ 0.3 \\
      & 35 & 5.41 & 10 $\pm$ 0.2 \\
\hline
\multirow{6}{*}{0.125} & 10 & -- & 9.1 $\pm$ 0.1 \\
      & 13 & -- & 8 $\pm$ 0.1 \\
      & 21 & 2.31 & 6.1 $\pm$ 0.1 \\
      & 25 & 2.72 & 5.6 $\pm$ 0.2 \\
      & 30 & 3.05 & 5 $\pm$ 0.1 \\
      & 35 & 3.38 & 4.4 $\pm$ 0.1 \\
\hline
\multirow{6}{*}{0.08}  & 10 & -- & 4.4 $\pm$ 0.1 \\
      & 13 & -- & 4.1 $\pm$ 0.02 \\
      & 21 & 1.48 & 3.4 $\pm$ 0.02 \\
      & 25 & 1.74 & 3.1 $\pm$ 0.01 \\
      & 30 & 1.95 & 2.8 $\pm$ 0.01 \\
      & 35 & 2.16 & 2.5 $\pm$ 0.03 \\
\hline
\multirow{3}{*}{0.05}  & 25 & 1.09 & 1.9 $\pm$ 0.01 \\
      & 30 & 1.22 & 1.7 $\pm$ 0.01 \\
      & 35 & 1.35 & 1.6 $\pm$ 0.02 \\
\hline
\end{tabular}
\end{minipage}%
\begin{minipage}[t]{.33\linewidth}
\vspace{0pt}
\centering
\begin{tabular}{ l  l  c  c }
\hline
\multicolumn{4}{c}{T4 DNA}\\
\hline
$c$       & T           & $c/c^{*}$ & $\eta_{0}$ \\  [0.5ex]
\hline
\hline
\multirow{4}{*}{0.214} & 13 & 2.08 & 128.4 $\pm$ 0.1 \\
      & 25 & 10.82 & 95.7 $\pm$ 0.4 \\
      & 30 & 11.89 & 85.3 $\pm$ 0.4 \\
      & 35 & 13.76 & 75.6 $\pm$ 0.4 \\
\hline
\multirow{8}{*}{0.148} & 10 & -- & 66.7 $\pm$ 0.3 \\
      & 13 & 1.44 & 58.5 $\pm$ 0.4 \\
      & 15 & 3.08 & 55.9 $\pm$ 0.7 \\
      & 18 & 4.93 & 50.7 $\pm$ 0.5 \\
      & 21 & 6.14 & 46.7 $\pm$ 0.5 \\
      & 25 & 7.48 & 42.5 $\pm$ 0.4 \\
      & 30 & 8.22 & 37.1 $\pm$ 0.2 \\
      & 35 & 9.52 & 32.5 $\pm$ 0.3 \\
\hline
\multirow{8}{*}{0.094} & 10 & -- & 21.9 $\pm$ 0.6 \\
      & 13 & 0.57 & 20.2 $\pm$ 0.8 \\
      & 15 & 1.96 & 19.2 $\pm$ 0.6 \\
      & 18 & 3.13 & 17.6 $\pm$ 0.4 \\
      & 21 & 3.92 & 16.6 $\pm$ 0.1 \\
      & 25 & 4.75 & 14.6 $\pm$ 0.3 \\
      & 30 & 5.22 & 12.9 $\pm$ 0.2 \\
      & 35 & 6.05 & 11.6 $\pm$ 0.2 \\
\hline
\multirow{6}{*}{0.059} & 15 & 1.23 & 10.2 $\pm$ 0.2 \\
      & 18 & 1.97 & 9.6 $\pm$ 0.1 \\
      & 21 & 2.46 & 8.9 $\pm$ 0.1 \\
      & 25 & 2.98 & 8.1 $\pm$ 0.03 \\
      & 30 & 3.28 & 7.3 $\pm$ 0.1 \\
      & 35 & 3.79 & 6.6 $\pm$ 0.1 \\
\hline
\multirow{5}{*}{0.038} & 18 & 1.27 & 4.9 $\pm$ 0.2 \\
      & 21 & 1.58 & 4.6 $\pm$ 0.2 \\
      & 25 & 1.92 & 4.2 $\pm$ 0.2 \\
      & 30 & 2.11 & 3.8 $\pm$ 0.2 \\
      & 35 & 2.44 & 3.3 $\pm$ 0.05 \\
\hline
\multirow{2}{*}{0.023} & 22 & 1 & 2.1 $\pm$ 0.01 \\
      & 24.5 & 1.1 & 2 $\pm$ 0.01 \\
\hline
\end{tabular}
\end{minipage}%
\end{table*}
%\endgroup

\subsection{\label{sec:thetascaling} Power law scaling at the $\theta$-temperature}

\begin{figure}[t]
\centering 
\subfloat[][]{ 
    \includegraphics[width=0.7\textwidth]{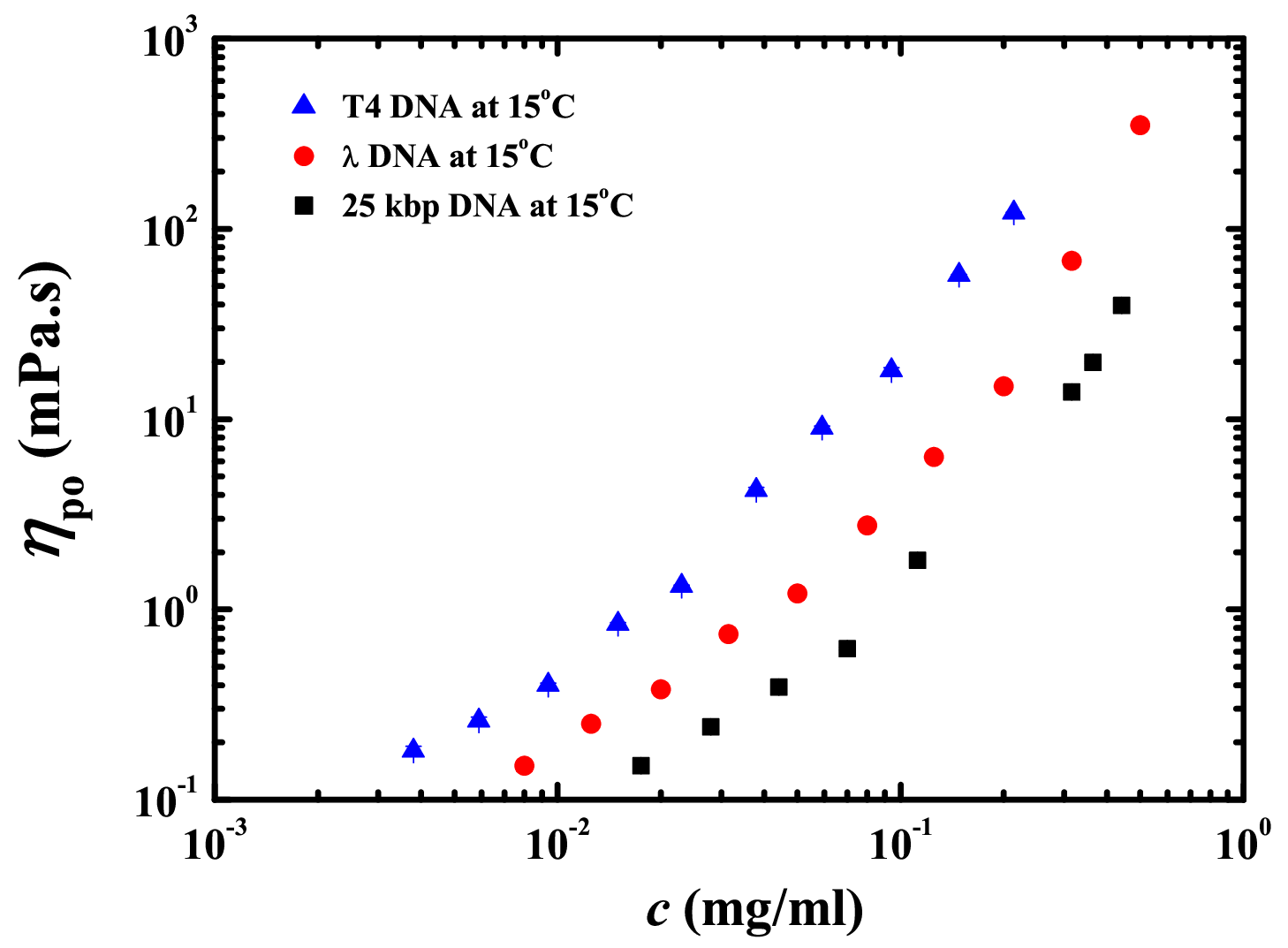}
  \label{figviscc:sub:a}}    \\
\subfloat[][]{ 
    \includegraphics[width=0.7\textwidth]{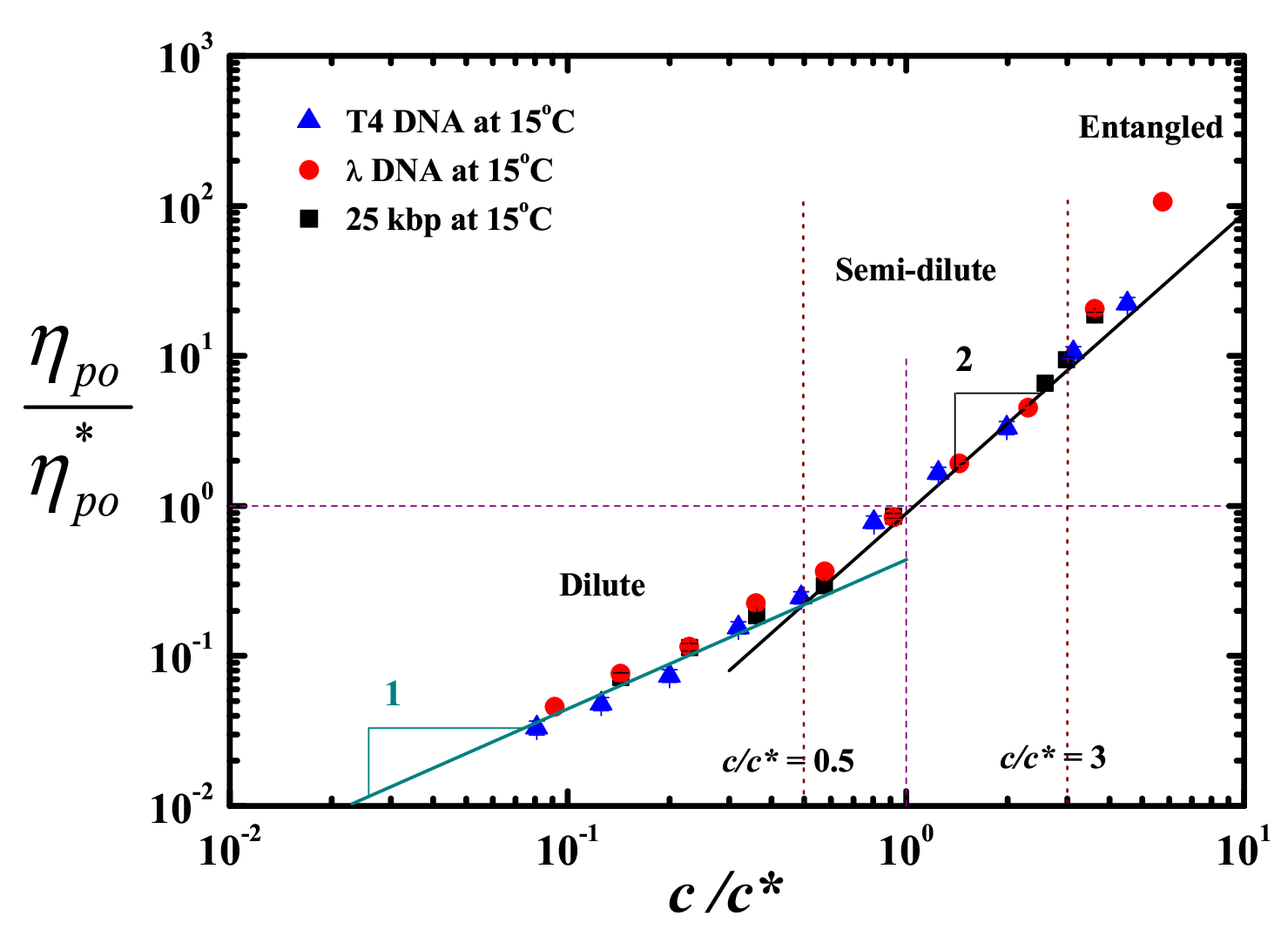}
   \label{figviscc:sub:b}}
%\begin{spacing}{1.5}
\caption{\label{fig:thetavisc} Dependence of the viscosity ratio $\eta_{p0}/\eta_{p0}^{*}$ (where $\eta_{p0}^{*}$ is the value of $\eta_{p0}$ at $c = c^{*}$) on the scaled concentration $c/c^{*}$, for 25 kbp, $\lambda$ and T4 DNA, at the $\theta$-temperature.}
%\end{spacing}
\end{figure}

Under $\theta$ solvent conditions, the polymer contribution to the zero shear rate viscosity is expected to obey the following scaling law in the semidilute unentangled regime~\citep{Jain2012a},
\begin{equation}
\frac{\eta_{p0}}{\eta_{p0}^{*}} \sim \left(\frac{c}{c^*}\right)^{2}
\end{equation}
where, $\eta_{p0}^{*}$ is the value of $\eta_{p0}$ at $c = c^{*}$. \citet{Jain2012a} have shown that it is more convenient to used $\eta_{p0}^{*}$ rather than $\eta_{s}$ as the normalising variable in the development of some of their scaling arguments. Additionally, it ensures that the ratio ${\eta_{p0}}/{\eta_{p0}^{*}} = 1$ when $c/c^{*} = 1$, for all the systems studied here. Clearly, when the bare zero shear rate viscosity versus concentration data [displayed in Fig.~\ref{fig:thetavisc}~(a)], is replotted in terms of scaled variables in Fig.~\ref{fig:thetavisc}~(b), data for the different molecular weight DNA collapse on top of each other, with the viscosity ratio depending linearly on $c/c^{*}$ in the dilute regime, followed by the expected power law scaling (with an exponent of 2) in the semidilute regime. Note that values of viscosity at  $T_\theta = 15$\degC, displayed in Fig.~\ref{fig:thetavisc}, were obtained by interpolation from values at nearby $T$ reported in Table~\ref{tab:allvisc}.

The semidilute unentangled regime is typically expected to span the
range from $c/c^{*} = 1$ to 10
\citep{Graessley1980,RubCol03}. Fig.~\ref{fig:thetavisc}~(b) suggests
that for $\theta$-solutions, the onset of the semidilute regime for the viscosity ratio, which
is dynamic property that is influenced by the presence of hydrodynamic
interactions, occurs with a relatively small crossover at a
concentration slightly less than $c/c^{*} = 1$.  Further, T4 DNA, which is the
longest molecule in the series studied here, appears to follow the
semidilute unentangled scaling for the largest concentration range,
while the 25 kbp and $\lambda$-phage DNA crossover into the entangled
regime beyond a concentration $c/c^{*} \gtrsim 3$. The difference in
the behaviour of the different DNA can be understood by the following
qualitative argument.

Chain entanglement is likely to occur when monomers from different
chains interact with each other. In a semidilute solution, this would
require a monomer within a concentration blob of one chain
encountering a monomer within the concentration blob of another
chain. A simple scaling argument suggests that at a fixed value of
$c/c^{*}$, such encounters become less likely as the molecular weight
of the chains increases. For a fixed value of $c/c^{*}$, it can be
shown that the number of concentration blobs in a chain remains
constant, independent of the molecular weight of the
chain~\citep{Jain2012a}. As a result, the size of a concentration blob
increases with increasing molecular weight, while at the same time the
concentration of monomers within a blob reduces. This decreasing
concentration within a blob makes entanglements less likely to occur
in systems with longer chains compared to systems with shorter chains,
at the \emph{same value} of $c/c^{*}$. This can also be seen from the
fact that, since in a semidilute solution the concentration within a
blob $c_\text{blob}$ is the same as the overall solution concentration
$c$, we can write $c_\text{blob} = (c/c^{*}) \times c^{*} \sim
(c/c^{*}) \, M^{1-3 \nu}$.

The scaling of the zero shear rate viscosity in semidilute solutions
under $\theta$ solvent conditions, displayed in
Fig.~\ref{fig:thetavisc}~(b), has been observed
previously~\citep{RubCol03}. However, to our knowledge, there have been
very few explorations in the experimental literature of the scaling of the
zero shear rate viscosity in the crossover region above the
$\theta$-temperature~\citep{Berry1996}. The experimental results we have obtained in
this regime are discussed within the framework of scaling theory in
the section below.

\subsection{\label{sec:crossoverscaling} Power law scaling in the crossover regime}

\begin{figure}[tbp]  \centering
\includegraphics[width=0.7\textwidth]{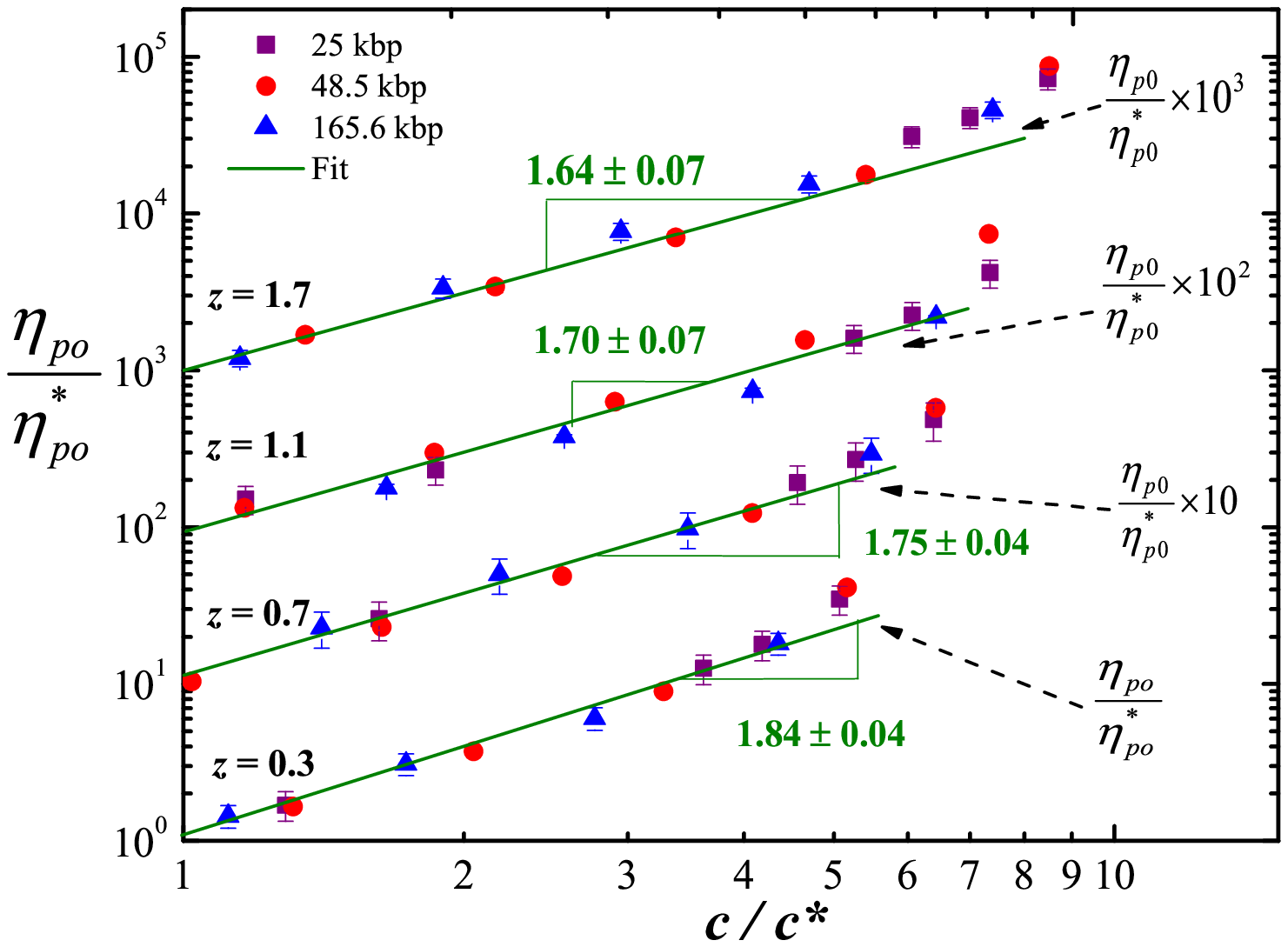}
\caption{Dependence of the viscosity ratio $\eta_{p0}/\eta_{p0}^{*}$
  on the scaled concentration $c/c^{*}$ in the semidilute regime, for
  25 kbp, $\lambda$ and T4 DNA, at fixed values of the solvent quality
  $z$. In order to display all the measurements on a single plot, viscosity ratios for the different values of $z$ have been multiplied by different fixed factors as indicated. Lines through the data are fits to the experimental data, with slopes and error in the fitted slope as shown. }
\label{fig:semidilutevis}
\end{figure}

The concentration dependence of the scaled polymer contribution to the
viscosity in the semidilute regime, ${\eta_{p0}}/{\eta_{p0}^{*}}$, for three different molecular
weights of DNA, is presented in Figure~\ref{fig:semidilutevis}, for
four different values of the solvent quality $z$. In order to maintain
the same value of solvent quality across the various molecular
weights, it is necessary to carry out experiments at the
appropriate temperature for each molecular weight. The relevant 
values of temperature at each value of $M$ are listed in 
Table~\ref{tab:nueffdil}. This procedure
would not be possible without the systematic characterisation of
solvent quality. Remarkably, Figure~\ref{fig:semidilutevis} indicates 
that, provided $z$ is the same, the data collapses onto universal power laws, independent of 
DNA molecular weight. Also worth noting is that while the crossover 
into the entangled regime for $\theta$-solutions occurs at around $c/c^{*} =3$, 
Figure~\ref{fig:semidilutevis} appears to suggest that the threshold for the onset of 
entanglement effects increases with increasing $z$.

As discussed earlier in section~\ref{sec:intro}, recent scaling theory
and Brownian dynamics simulations~\citep{Jain2012a} suggest that the
viscosity ratio should scale according to the power law,
\begin{equation}
\label{eq:eta0crossover}
\frac{\eta_{p0}}{\eta_{p0}^{*}} \sim \left(\frac{c}{c^*}\right)^{{ 1
  }/({3 \nueff(z) -1 })} 
\end{equation}
where, the dependence of the effective exponent $\nueff$ on the
solvent quality $z$ should be identical to that which characterises
the power laws for both the polymer size and the diffusivity. From the
set of values of $z$ for which Brownian dynamics simulations results
have been reported by~\citet{Jain2012a}, there are two values at which
this conclusion can be tested by comparison with experiment, namely, $z = 0.7$
and $z=1.7$. (Note that at each value of $z$, the experimental value of $\nueff$ can be determined by equating the slope of the fitted lines in Figure~\ref{fig:semidilutevis} to ${1}/({3 \nueff -1 })$). The values of $\nueff(z)$ listed in Table~\ref{tab:nueffdil}, at $z =0.7$ and $1.7$, suggest that simulation and experimental exponents agree with each other to within error bars.
\begin{table}[t]
\caption{Values of the effective exponent $\nu_{\mathrm{eff}}(z)$ determined experimentally at $z = \{0.3, 0.7, 1.1, 1.7\}$ and by Brownian dynamics simulations at $z = \{0.7, 1.7\}$.}
\vskip20pt
\begin{tabular}{c  c  c  c  c  c  c}
\hline
$z$ & 25 kbp & $\lambda$-DNA & T4 DNA & $\partial \ln(\eta_{p0}/\eta_{p0}^{*})/\partial \ln(c/c^{*})$ & $\nu_{\mathrm{eff}}$ & $\nu_{\mathrm{eff}}$ \\
 & $T$ & $T$ & $T$ & (experiments) & (experiments) & (BDS)\\
\hline
\hline
0.3 & 19.7\degC & 18.4\degC & 16.8\degC & 1.84 $\pm$ 0.04 & 0.51 $\pm$ 0.01 & -- \\
0.7 & 26.1\degC & 22.9\degC & 19.2\degC & 1.75 $\pm$ 0.04 & 0.52 $\pm$ 0.01 & 0.54 $\pm$ 0.02 \\
1.1 & 32.8\degC & 27.5\degC & 21.7\degC & 1.70 $\pm$ 0.07 & 0.53 $\pm$ 0.01 & -- \\
1.7 & 43.4\degC & 34.8\degC & 25.4\degC & 1.64 $\pm$ 0.07 & 0.54 $\pm$ 0.01 & 0.58 $\pm$ 0.03 \\
%3   & --              & --              & 0.63 $\pm$ 0.03 \\
\hline
\end{tabular}
\label{tab:nueffdil}
\end{table}

\subsection{\label{sec:stein} Universal ratio of relaxation times}

Blob scaling arguments can be used to show that, away from the crossover boundaries, the concentration dependence of the longest relaxation time $\lambda_{1}$, obeys the power law,
\begin{equation}
\lambda_{1} \sim  \left(\frac{c}{c^*}\right)^{ ({2-3 \nu })/({3 \nu - 1})}
\label{lambda1}
\end{equation}
In very good solvents, since $\nu \approx 0.59$, this would imply $\lambda_{1} \sim  \left({c}/{c^*}\right)^{0.3}$, while in $\theta$-solutions, $\lambda_{1} \sim  {c}/{c^*}$. 

\citet{Liu20091069} have recently examined the concentration dependence of $\lambda_{1}$ by studying the relaxation of stretched single T4 DNA molecules in semidilute solutions. They find that at $22 \degC$, the longest  relaxation time obeys the power law,
\begin{equation}
\frac{\lambda_{1}}{\lambda_{1,z}} \sim  \left(\frac{c}{c^*}\right)^{0.5}
\end{equation}
where, $\lambda_{1,z}$ is the longest relaxation time in the dilute
limit. This clearly suggests that, (i) for the solution of T4 DNA
considered in their work, $22 \degC$ is in the crossover regime, and
(ii) the relaxation time also obeys a power law in the crossover
regime (as observed here for viscosity), with an effective exponent $\nueff \approx 0.56$.  

It is worth noting that, for T4 DNA molecules dissolved in the solvent used in the present work, $22 \degC$ corresponds to a value of the solvent quality parameter $z=1.17$.
\begin{figure}[t]  \centering
\includegraphics[width=0.75\textwidth]{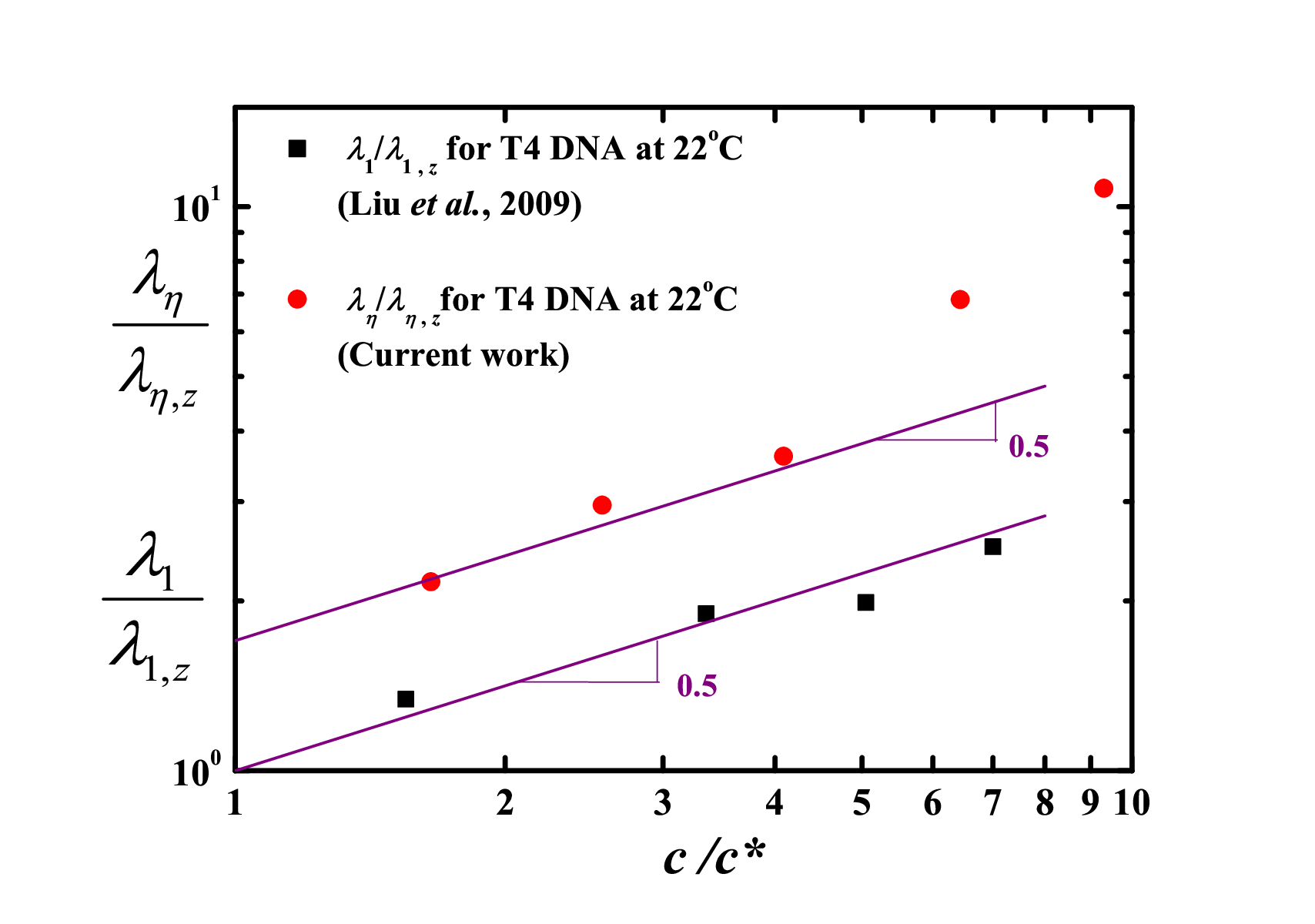}
\caption{The concentration dependence of the ratio ${\lambda_{1}}/{\lambda_{1,z}}$  in the semidilute regime, obtained by \citet{Liu20091069}, compared with the dependence of the ratio ${\lambda_{\eta}}/{\lambda_{\eta,z}}$, measured by current experiments at $22 \degC$.}
\label{fig:relaxcomp}
\end{figure}

It is common to define an alternative large scale relaxation time
$\lambda_{\eta}$, based on the polymer contribution to the zero shear rate viscosity $\eta_{p0}$, by the following expression~\citep{ottinger},
\begin{equation}
\lambda_\eta = \frac{M \eta_{p0}}{c N_A k_B T}
\label{eq:lameta}
\end{equation}
where, $k_{B}$ is Boltzmann's constant. It is straight forward to show that, in the semidilute unentangled regime, $\lambda_{\eta}$ obeys the same power law scaling with concentration as obeyed by $\lambda_{1}$ [see Eq.~(\ref{lambda1})]. Fig.~\ref{fig:relaxcomp} compares the concentration dependence of the ratio ${\lambda_{1}}/{\lambda_{1,z}}$  in the semidilute regime, obtained by \citet{Liu20091069}, with that of the ratio ${\lambda_{\eta}}/{\lambda_{\eta,z}}$, measured by the current experiments at $22 \degC$. Here, $\lambda_{\eta,z}$ is a large scale relaxation time in the dilute limit, defined by the expression $\lambda_{\eta,z} = M [\eta]_{0}\eta_{s}/ N_A k_B T$, where $[\eta]_{0}$ is the zero shear rate intrinsic viscosity. It is clear that both relaxation times exhibit identical scaling with concentration in the semidilute regime at $22 \degC$. 

It is well known that for dilute polymer solutions, the ratio of the two large scale relaxation times, 
\begin{equation}
U_{\eta \lambda} =\frac{\lambda_{\eta,z}}{\lambda_{1,z}} 
\label{uetalambda}
\end{equation}
is a universal constant, independent of polymer and solvent chemistry. Predicted values of $U_{\eta \lambda}$ vary from 1.645 by Rouse theory to 2.39 by Zimm theory, with predictions by other approximate theories lying somewhere in between~\citep{Kroger20004767}. Recently, \citet{somani2010} have predicted the dependence of $U_{\eta \lambda}$ on the solvent quality $z$, in the dilute limit, with the help of Brownian dynamics simulations. This enables us to calculate the value of the ratio $\lambda_{\eta}/\lambda_{1}$ at $22 \degC$ using the present measurements and the measurements of \citet{Liu20091069}, by the following argument. Clearly, 
\begin{equation}
\frac{\lambda_{\eta}}{\lambda_{1}}   =  \left( \frac{\lambda_{\eta}}{\lambda_{\eta,z}} \right)  \left( \frac{\lambda_{1,z}}{\lambda_{1}} \right)  U_{\eta \lambda} (z)
\label{etalambda}
\end{equation}
Since the effective exponent in the experiments of  \citet{Liu20091069} and the present experiments is the same ($\nueff = 0.56$), we assume that the two solutions have the same value of $z=1.17$. At this value of $z$, the simulations of \citet{somani2010} suggest that $U_{\eta \lambda} \, (z = 1.17) = 1.79$. Equation~(\ref{etalambda}) can then be used to find the ratio ${\lambda_{\eta}}/{\lambda_{1}}$ at the various values of concentration at which the ratios $\lambda_{\eta}/\lambda_{\eta,z}$ and $\lambda_{1}/\lambda_{1,z}$ have been measured in the two sets of experiments. 

\begin{figure}[tbp]  \centering
\includegraphics[width=0.75\textwidth]{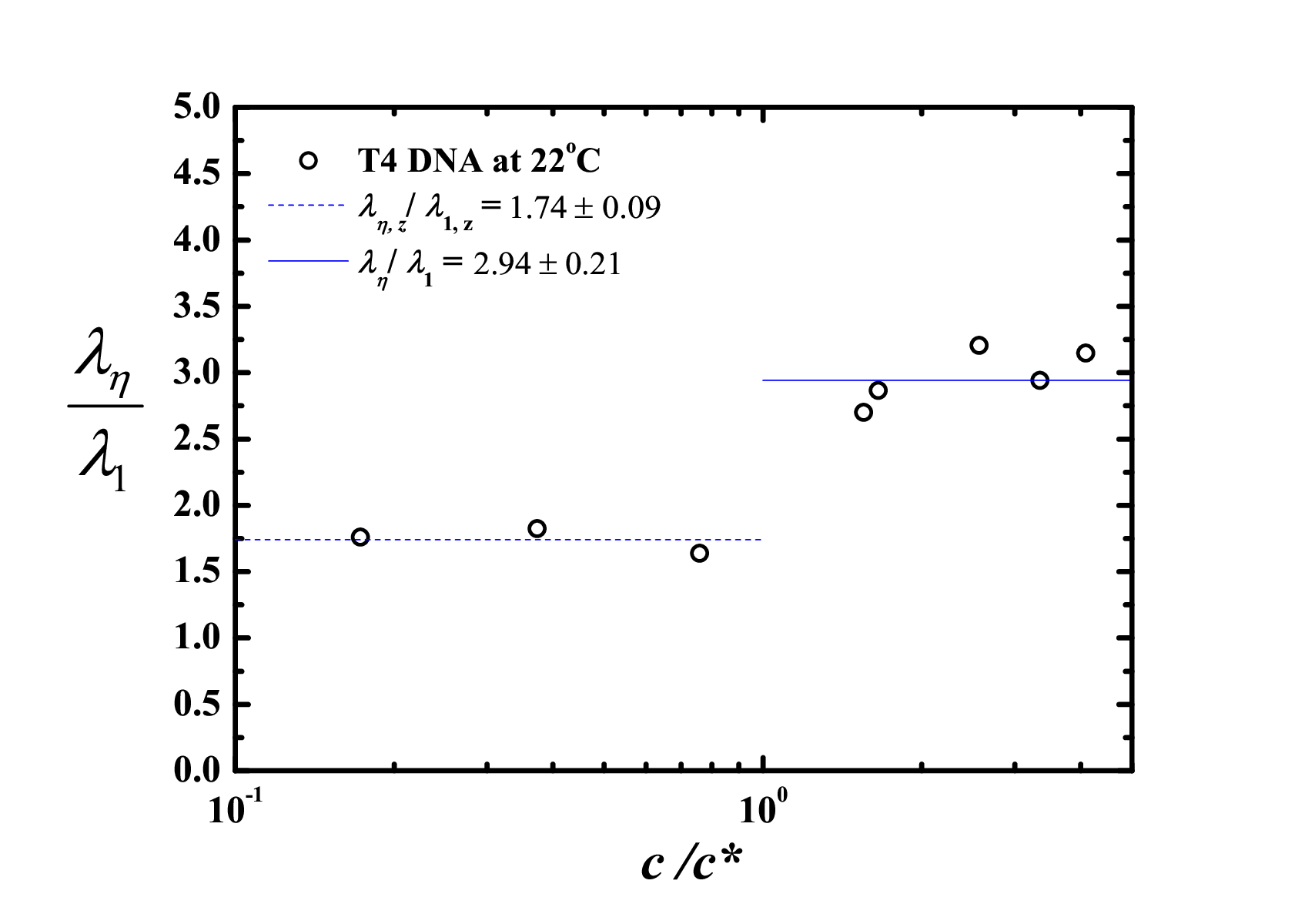}
\caption{Universal ratio of $\lambda_{\eta}$ (measured in the present work) to $\lambda_{1}$ (measured by \cite{Liu20091069}) for T4 DNA at $22 \degC$ (which corresponds to a value of solvent quality $z=1.17$), for a range of concentrations spanning the dilute and semidilute regime.}
\label{fig:relaxratio}
\end{figure}

Figure~\ref{fig:relaxratio} displays the ratio $\lambda_{\eta}/\lambda_{1}$ obtained in this manner in the dilute and semidilute regimes. Since both the ratios $\lambda_{\eta}/\lambda_{\eta,z}$ and $\lambda_{1}/\lambda_{1,z}$  are nearly equal to 1 in the limit of small $c$, it is not surprising that $\lambda_{\eta}/\lambda_{1} \approx U_{\eta \lambda} \, (z)$, for concentrations in the dilute regime. However, while $\lambda_{\eta}/\lambda_{1}$ is constant in the semidilute regime, as expected from the similar scaling with concentration exhibited in the two sets of experiments, its value is not identical to the value in the dilute limit. This appears to be because $\lambda_{\eta}/\lambda_{\eta,z}$ increases more rapidly with concentration in the crossover regime between dilute and semidilute, than  $\lambda_{1}/\lambda_{1,z}$. More experiments carried out for different polymer solvent systems are required to substantiate this observation.

\section{\label{sec:con} Conclusions}

By carrying out accurate measurements of the polymer contribution to the zero-shear rate viscosity of semidilute DNA solutions in the double crossover regime, the scaled polymer contribution to the viscosity is shown to obey the expression,
\begin{equation*}
\frac{\eta_{p0}}{\eta_{p0}^{*}} \sim \left(\dfrac{c}{c^*}\right)^{\tfrac{ 1
  }{3 \nueff(z) -1} }
\end{equation*}
in line with recent predictions on the form of universal crossover scaling functions for semidilute solutions~\citep{Jain2012a}. The experimentally determined values of the effective exponent $\nueff$, for two values of $z = \{0.7, 1.7\}$, agree within error bars, with values determined from Brownian dynamics simulations. This suggests, in accordance with the prediction of scaling theory~\citep{Jain2012a},  that the exponent $\nueff (z)$ that governs the scaling of viscosity is identical to the exponent which characterises the power laws for polymer size and the diffusivity. 

The demonstration of this scaling behaviour requires the determination of the $\theta$ temperature of the model DNA solutions used here, and a characterisation of its solvent quality. By carrying out static  light scattering measurements, the $\theta$ temperature of the aqueous dilute solution of DNA (in excess sodium salt) has been determined to be $T_\theta = 14.7 \pm 0.5$\degC, while dynamic light scattering measurements have been performed to find the solvent quality of the DNA solutions, at any given molecular weight $M$ and temperature $T$. 

The results obtained here clearly demonstrate that the solvent quality parameter $z$, and the scaled concentration $c/c^{*}$, are the two scaling variables that are essential in order to properly understand and characterise the concentration and temperature dependent dynamics of a linear viscoelastic property, such as the zero shear rate viscosity, of semidilute polymer solutions. These results are also relevant to obtaining a universal description of polymer solution behaviour \emph{away from equilibrium}, since it would be necessary to specify the values of $z$, $c/c^{*}$, and the Weissenberg number $W\!i$ (which is the scaling variable that characterises flow), in order to obtain a complete description of the state of the solution.

%%%%%%%%%%%%%%%%%%%%%%%%%%%%%%%%%%%%%%%%%%%%%%%%%%%%%%%%%%%
\begin{acknowledgments}
  This research was supported under Australian Research Council's Discovery Projects funding scheme (project number DP120101322). We are grateful to D. E. Smith and his group (University of
  California, San Diego, USA) for preparing the majority of the
  special DNA fragments and to B. Olsen (MIT, Cambridge, USA) for
  the stab cultures containing them. We thank S. Noronha and his
  group (IIT Bombay, India) for helpful discussions, laboratory
  support, and the two plasmids, pBSKS and pHCMC05, used in this
  work. We are greatly indebted to S. Bhat and K. Guruswamy (National Chemical
  Laboratory, Pune, India), for assistance with the light scattering experiments, 
  and for helpful discussions in this regard. JRP gratefully acknowledges
  extensive and very helpful discussions on numerous topics in this paper 
  with B. D\"unweg (MPIP, Mainz, Germany). We
  also acknowledge the equipment support received through DST (IRHPA:
  Liposomes) and consumables through MHRD (IITB) funds.  The Sophisticated
  Analytical Instrument Facility (SAIF), IIT Bombay is thanked for
  access to the static light scattering facility. 
\end{acknowledgments}
\vspace{12pt}

See supplementary material at [URL to be inserted by AIP] for details
of strains and working conditions; procedures for preparation of
linear DNA fragments; solvent composition and quantification of DNA
samples; procedure for sample preparation for light scattering, and
the methodology for estimating the second virial coefficient and the
hydrodynamic radius from the light scattering measurements.

\appendix

\section{\label{sec:sls} Determining the $\theta$-temperature of the DNA solutions}

\begin{figure}[!bt]
\begin{center}
\epsfig{file=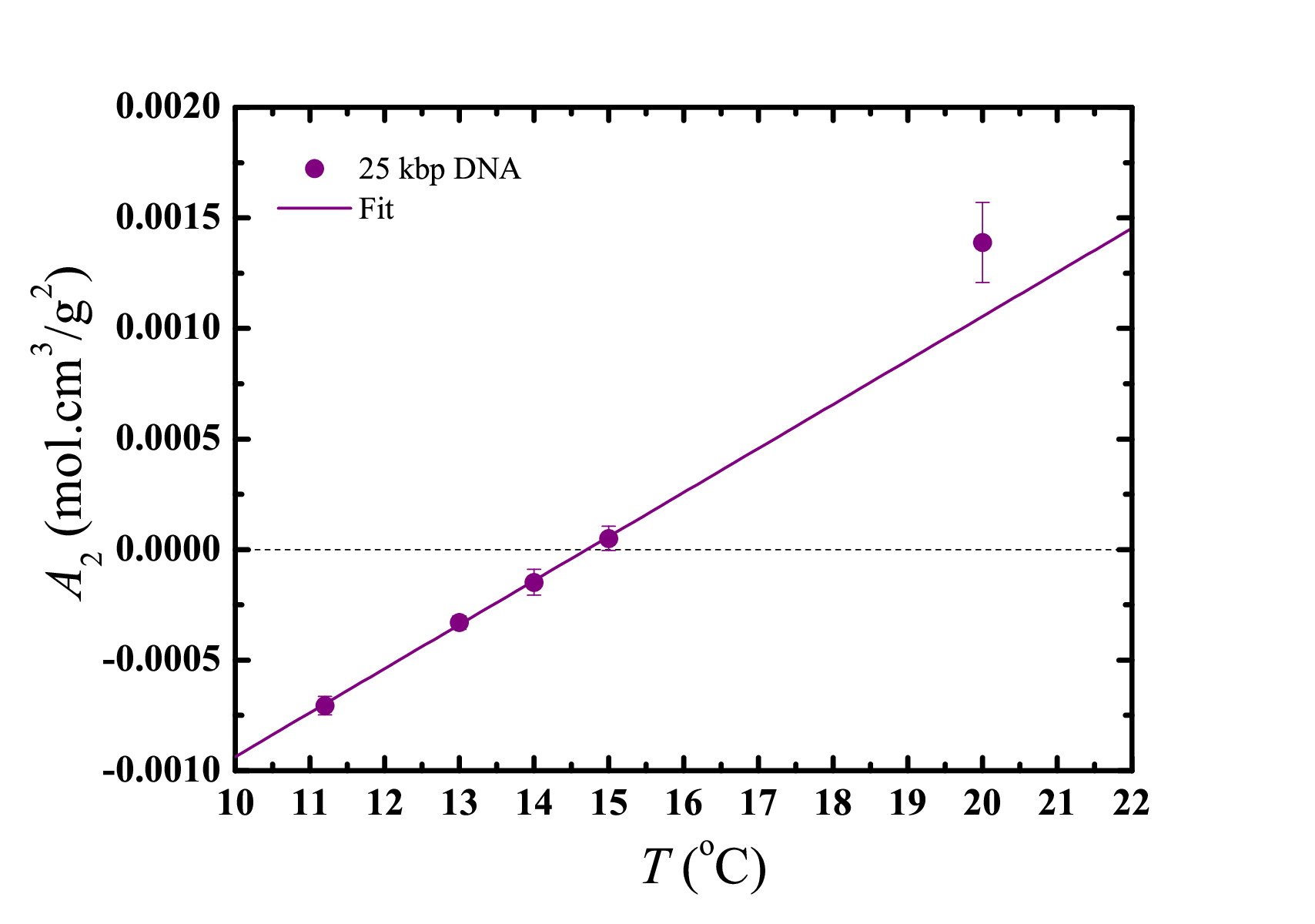,width=0.7\linewidth,clip=}
\end{center}
\caption{Determination of the $\theta$ temperature, $T_{\theta}$, for 25 kbp DNA. The equation of the fitted line to the temperature dependence of the second virial coefficient is: $A_{2} = -3.15 \times 10^{-3}+ 2.16 \times 10^{-4} \, T$, where $T$ is in \degC.}
\label{fig:A2}
\end{figure}

\begin{figure}[t]
\begin{center}
\epsfig{file=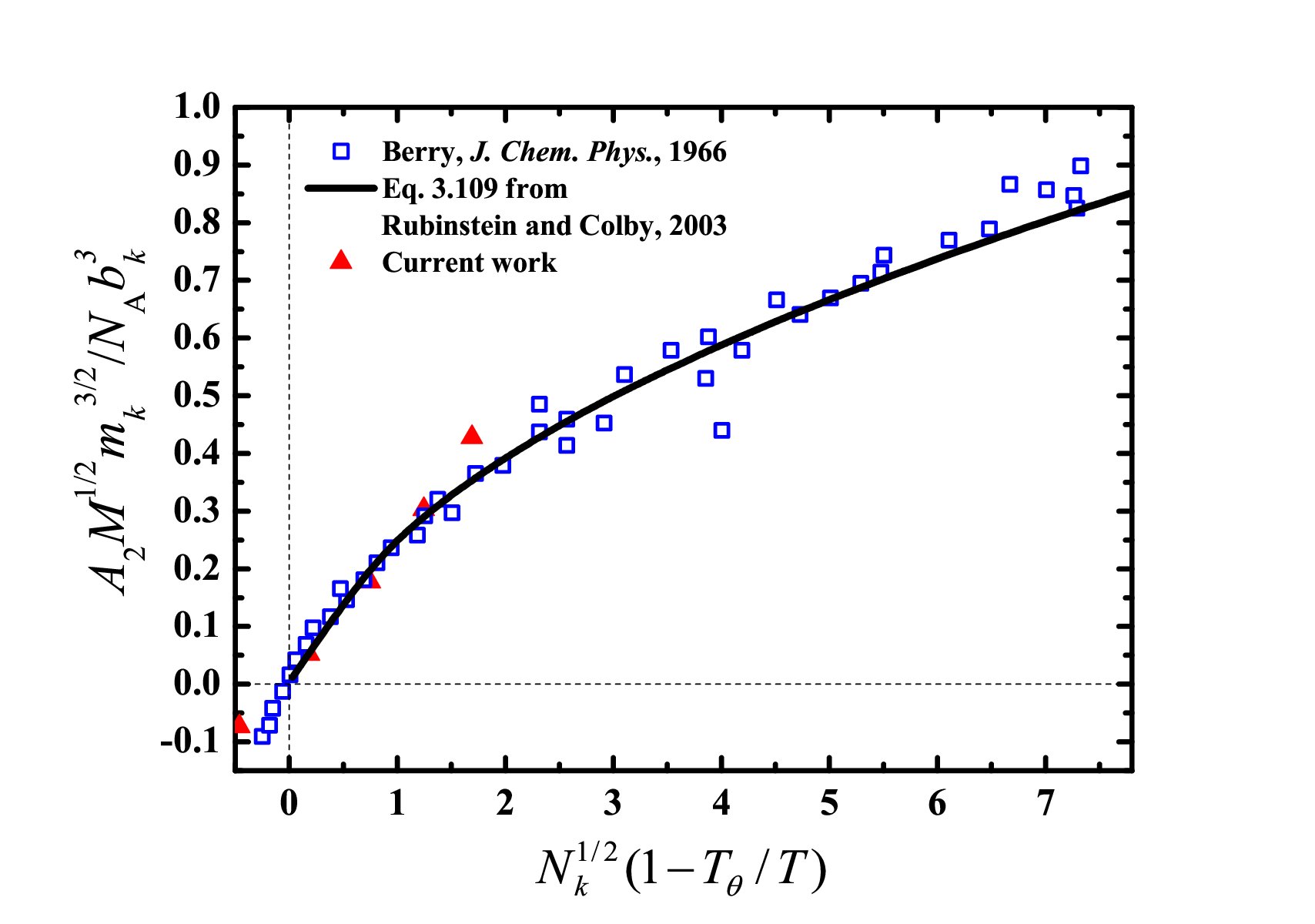,width=0.7\linewidth,clip=}
\end{center}
\caption{Universal crossover plot for the second virial
  coefficient. Values of $A_{2}$ for 25 kbp DNA (red triangles) are
  calculated from the fit function given in the caption to
  Fig.~\ref{fig:A2} at 14, 15, 16, 17 and 18\degC. The $\theta$
  temperature is taken to be $T_{\theta} = 14.7\degC$. The line is
  drawn according to Eq.~(\ref{A2MP}). The molar mass per Kuhn step is
  defined as $m_{k} = M / N_{k}$, and the Kuhn step length is $b_{k} =
  2 P$. Values of $M$, $N_{k}$, and $P$ are given in
  Table~\ref{tab:relaxationtime}. Open squares represent data from
  \citet{berry1966}, for polystyrene in decalin.}
\label{fig:A2mp}
\end{figure}

\begin{figure}[t]
\begin{center}
\epsfig{file=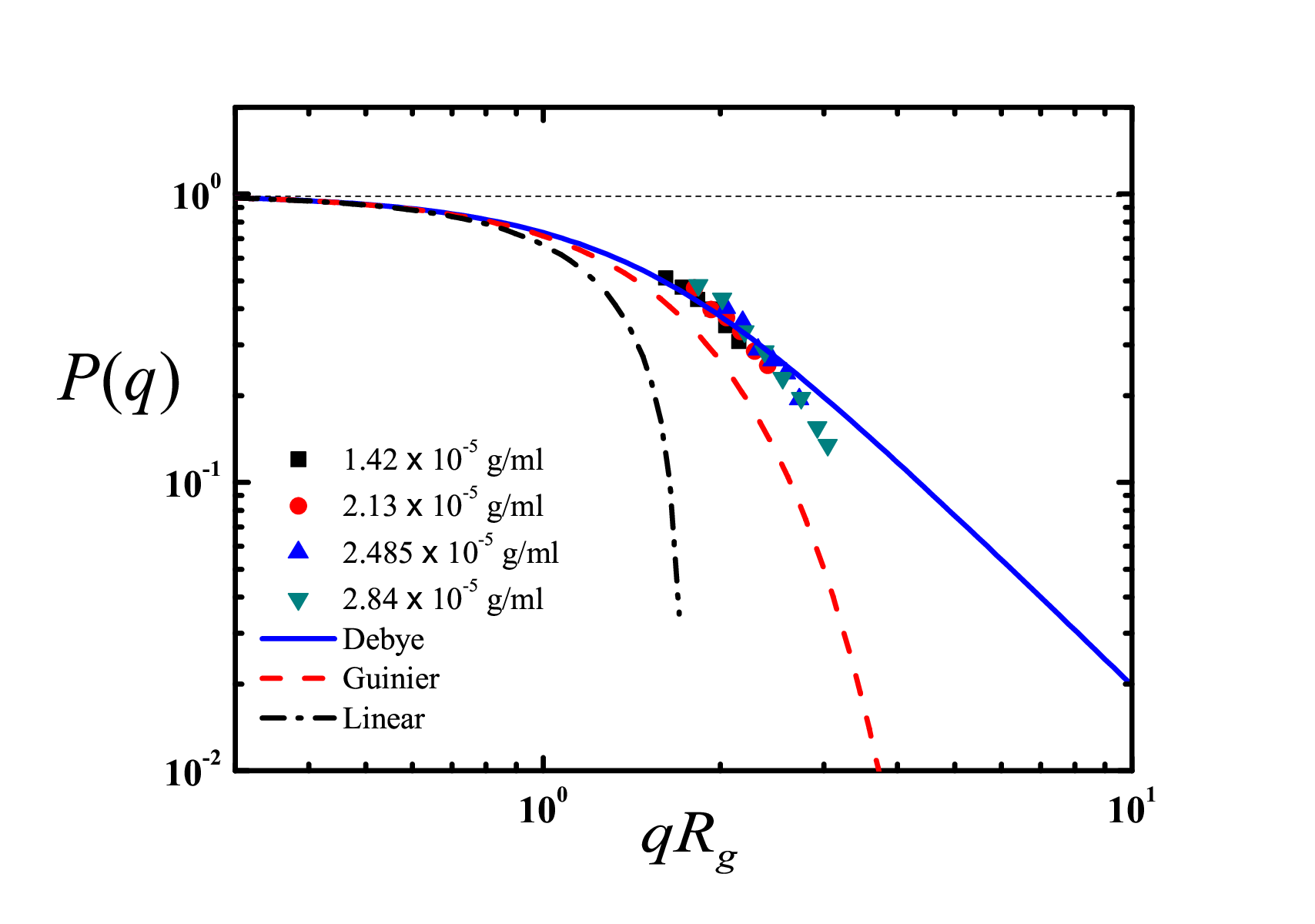,width=0.7\linewidth,clip=}
\end{center}
\caption{A Debye function fit to the form factor, $P(q)$, data for 25 kbp DNA, obtained at $14 \degC$ and four different concentrations. The Guinier approximation, $P(q) = \exp \left( - {q^{2}{\Rg}^{2}}/{3}  \right)$, and the linear approximation, $P(q) = 1- \left(q^{2}{\Rg}^{2}/3 \right)$ are also displayed.}
\label{fig:Debye25kbp}
\end{figure}

The $\theta$ temperature for a polymer solution can be determined by finding the temperature at which the second virial coefficient $A_{2}$ is zero. One of the methods often used to determine the temperature dependence of $A_{2}$  is static light scattering, since the intensity of scattered light, $I(q)$, at any temperature, concentration and molecular weight of the dissolved species, depends on $A_{2} (T)$. Details of the static light scattering experiments, the governing equation for $I(q)$, and the procedure adopted here to determine $A_{2}(T)$, are discussed in the supplementary material. The principal results of the analysis are presented here. 

Figure~\ref{fig:A2}, which is a plot of the second virial coefficient for 25 kbp DNA as a function of temperature in the range 10 to 20 \degC, shows that $A_{2}$ increases from being below zero to above zero in this range of temperatures. A linear least squares fit to the data in the vicinity of the $\theta$ temperature (where the dependence is expected to be linear) suggests that, $T_{\theta} = 14.7 \pm 0.5 \degC$. Note that this implies that a significant fraction of the temperatures at which measurements were carried out are in the poor solvent regime. The reliability of the current measurements in the poor solvent regime is discussed in detail in Appendix~\ref{sec:blob}.

As in the case of other polymer solution properties, the second virial
coefficient, when represented in a suitably normalised form, is a
universal function of the solvent quality parameter in the crossover
region. The specific form of the crossover function used to describe
the dependence is,
\begin{equation}
\frac{A_{2} M^{\frac{1}{2}} m_{k}^{\frac{3}{2}}}{N_{A} b_{k}^{3}} = 0.20 \left[ {\tilde z}^{-2.64} + {\tilde z}^{-1.4} \right]^{-0.38}
\label{A2MP}
\end{equation}
where, $\tilde z = 2N_{k}^{1/2} \left(1- \dfrac{T_{\theta}}{T}\right)$ [see Eq.~(3.109) in \citet{RubCol03}]. The temperature and molecular weight dependence of the second virial coefficient, for a number of polymer-solvent combinations, and from computer simulations, is found to obey this universal crossover function. Figure~\ref{fig:A2mp} is a plot of this function, which is modelled after a similar figure in \citet{RubCol03}, along with the data reported previously by \citet{berry1966} for linear polystyrenes in decalin. We have used a linear least squares fit to the 25 kbp DNA data displayed in Fig.~\ref{fig:A2}, and evaluated $A_{2}$ at a few temperatures between 14 and 20~\degC\ (indicated by the red triangles in Fig.~\ref{fig:A2mp}). Clearly the present data also appears to lie on the universal crossover function. 

\begin{figure}[t]  \centering
\includegraphics[width=0.7\textwidth]{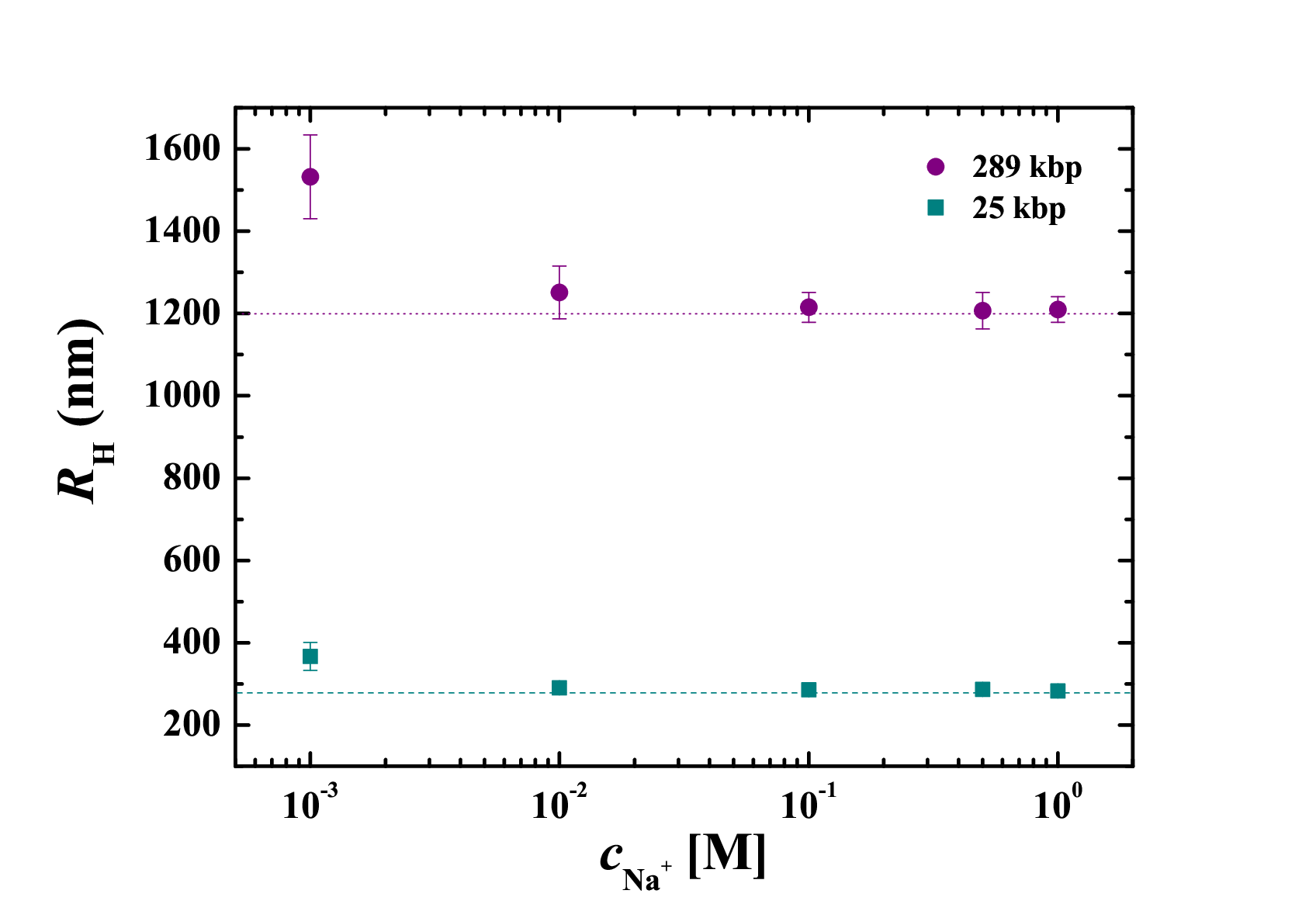}
\caption{Dependence of \Rh\ on salt concentration for two different molecular weights at 25 \degC.}
\label{fig:salt1}
\end{figure}

At the $\theta$ temperature, the precise form of the expression for the form factor, $P(q) = I(q)/I_{0}$, where, $I_{0} = \lim_{q \to 0} I(q)$, is known to have the following form (referred to as the Debye function~\citep{RubCol03}),
\begin{equation}
P(q) = \frac{2}{\left(q^{2} \, {\Rgth}^{2} \right)^{2}} \left[ \exp \left(-q^{2} \, {\Rgth}^{2} \right) - 1 + q^{2} \, {\Rgth}^{2} \right]
\label{debye}
\end{equation}
Note that, since we know the contour length and the persistence length for 25 kbp DNA, we can estimate $\Rgth = 376$ nm, as displayed in Table~\ref{tab:relaxationtime}. The determination of $I_{0}$ from the measured $I(q)$ data for 25 kbp DNA is discussed in the supplementary material. As a result, the dependence of $P(q)$ on $q$, for the current measurements on 25 kbp DNA, is known. The Debye function is also known to describe the angular dependence of the scattered intensity at temperatures away from the $\theta$ temperature very well, over a wide range of values of $q^{2}  {\Rg}^{2}$~\citep{Schafer99,Utiyama1971}. As a result, the Debye function can be used to fit the $P(q)$ data to determine \Rg. Figure~\ref{fig:Debye25kbp} displays the Debye function fit to the $P(q)$ data for 25 kbp DNA, at $14 \degC$ and four different concentrations, along with the Guinier approximation $P(q) = \exp \left( - {q^{2}{\Rg}^{2}}/{3}  \right)$, and the linear approximation, $P(q) = 1- \left(q^{2}{\Rg}^{2}/3 \right)$. We can see that the Debye function describes the data reasonably accurately, independent of concentration, over a wide range of the measured values of $q^{2}  {\Rg}^{2}$. We find that the fitted values of \Rg\ are in the range $389.4 \pm 68.1$ nm across the four different concentrations. While this is reasonably close to the analytical value of $\Rgth = 376$ nm, as is expected at $14 \degC$, the current scattering data does not cover a sufficiently wide range of $q^{2}  {\Rg}^{2}$ values to determine \Rg\ more precisely. 

\section{\label{sec:dls} Estimating the solvent quality of the DNA solutions}

The scaling variable that describes the temperature crossover behaviour from $\theta$ solvents to very good solvents,  is the solvent quality parameter $z$, defined by the expression~\citep{Schafer99},
 \begin{equation}
 \label{eq:z}
z = k\left(1- \dfrac{T_{\theta}}{T} \right) \, \sqrt{M}
\end{equation}
where, $k$ is a chemistry dependent constant that will be discussed in greater detail shortly below. The significance of the variable $z$ is that when data for any equilibrium property of a polymer-solvent system is plotted in terms of $z$ in the crossover region, then regardless of the individual values of $M$ and $T$, provided the value of $z$ is the same, the equilibrium property will turn out to have the same value. Indeed, provided the values of $k$ are chosen appropriately, equilibrium data for different polymer-solvent systems  can be shown to collapse onto master plots, revealing the universal nature of polymer solution behaviour. Typically, a particular polymer-solvent system is chosen as the reference system and data for all other systems are shifted to coincide with the values of the reference system by an appropriate choice of $k$~\citep{MiyFuj81,Tominaga20021381,Hayward19993502}. The same shifting procedure is also commonly used to compare experimental observations in the crossover regime with theoretical predictions or simulations results~\citep{Kumar20037842,SunRav06-epl}. Basically, as will be demonstrated in greater detail subsequently, the value of $k$ for an experimental system is chosen such that the experimental and theoretical values of $z$ agree when the respective equilibrium property values are identical. 

We have determined the value of $z$ for the DNA solutions used here by comparing experimental measurements of the swelling of the hydrodynamic radius $\alpha_{\mathrm{H}} =  {\Rh(T)}/ {\Rhth}$, where \Rh\ is the hydrodynamic radius, with predictions of Brownian dynamics simulations reported previously~\citep{SunRav06-epl}. The hydrodynamic radius has been measured by carrying out dynamic light scattering measurements over a range of temperatures and molecular weights at a concentration $c/c^{*}$ = 0.1. Details of the dynamic light scattering measurements, including the instrument used, sample preparation procedure, and typical intensity plots are given in the supplementary material. Before discussing the details of the estimation of solvent quality, it is appropriate to  first present some results of the measurements of the hydrodynamic radius. 

\begin{figure}[tbp]  \centering
\includegraphics[width=0.68\textwidth]{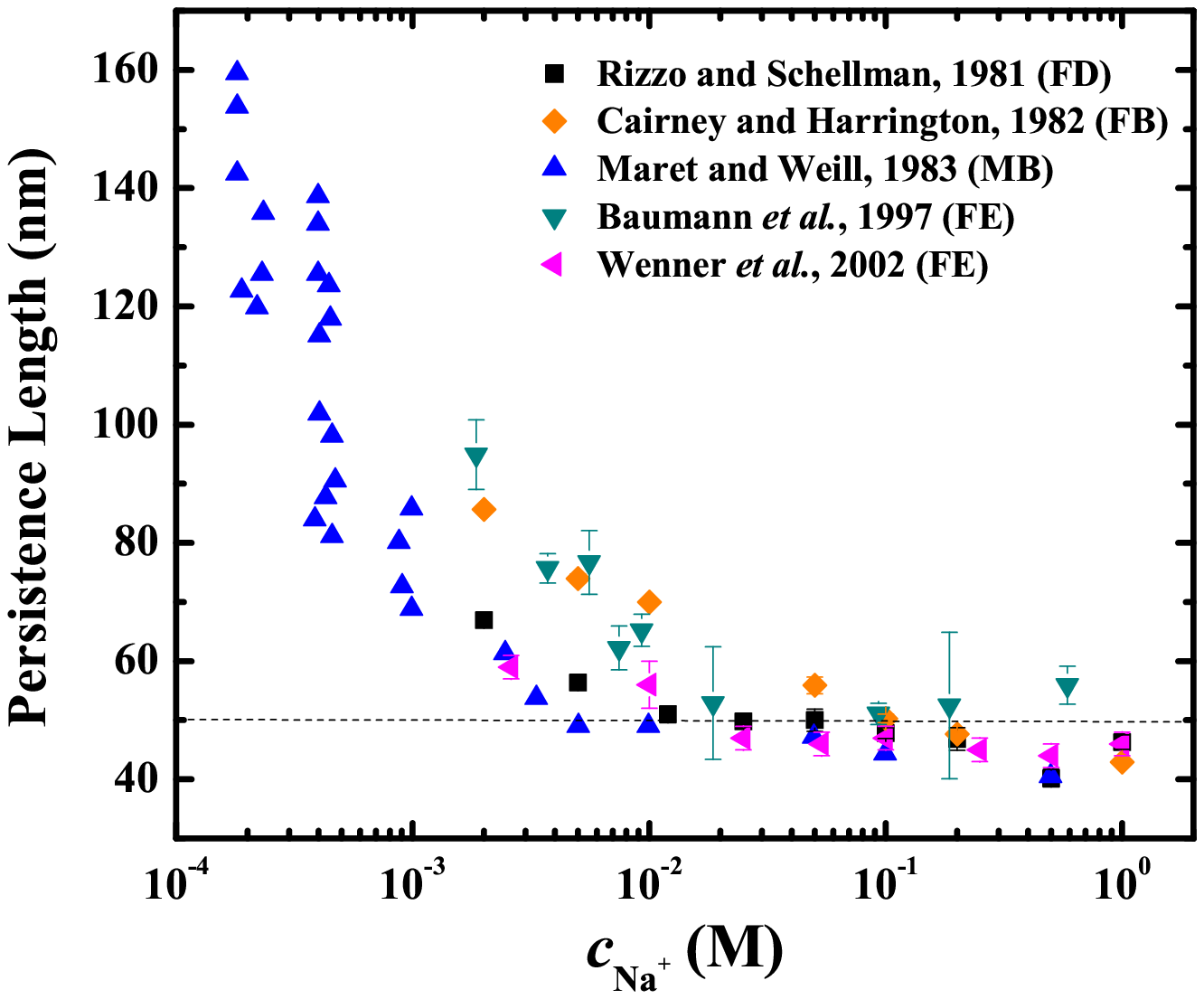}
\caption{Dependence of DNA persistence length on salt concentration, collated from data reported previously. The DNA molecular weights used in all these studies range from 3--300 kbp. Abbreviations: `FD' - Flow Dichroism, `FB' - Flow Birefringence, `MB' - Magnetic Birefringence, `FE' - Force Extension using optical tweezers.}
\label{fig:salt2}
\end{figure}

\begin{figure}[tbp]  \centering
\includegraphics[width=0.7\textwidth]{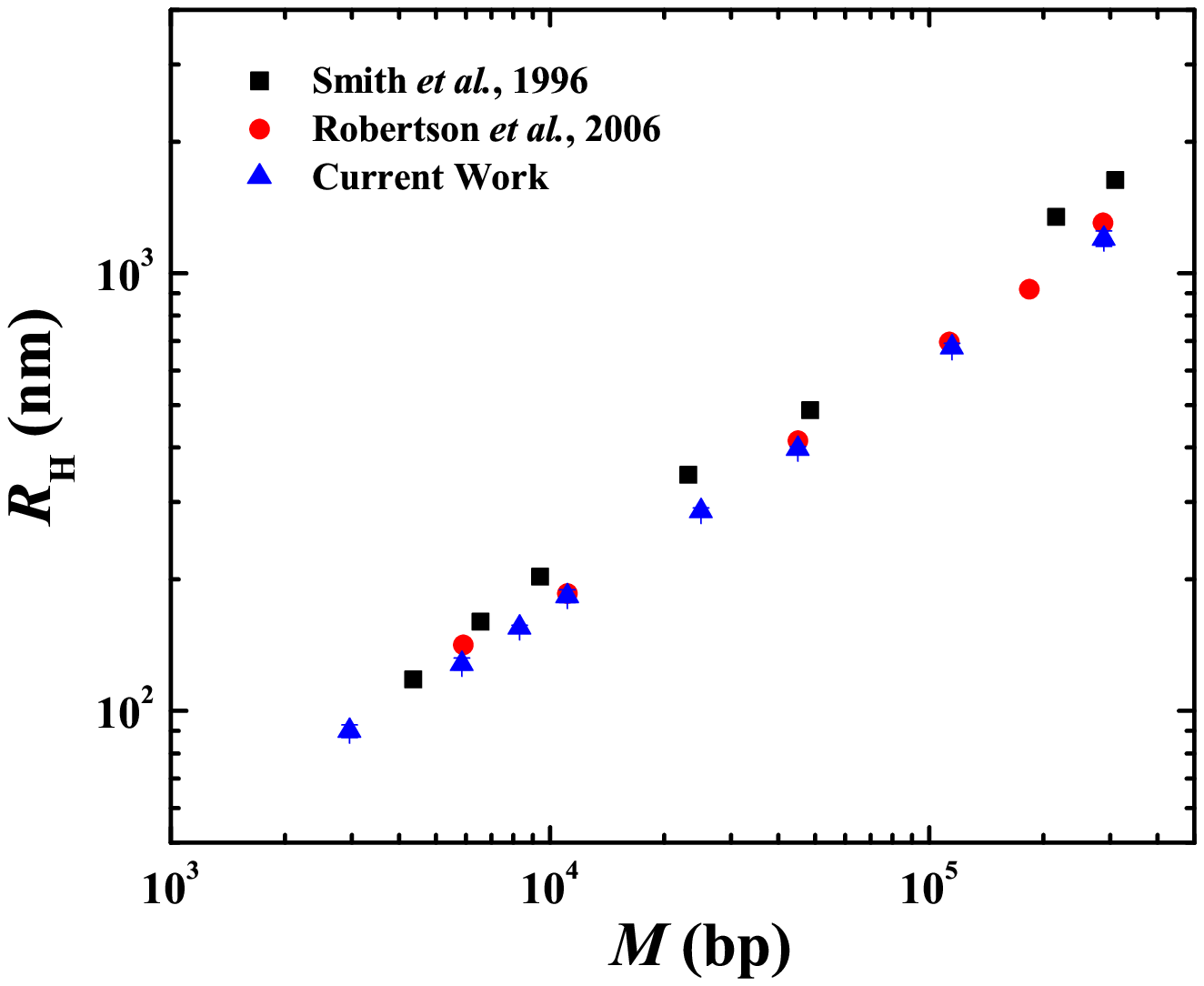}
\caption{Comparison of the molecular weight dependence of hydrodynamic radius, obtained previously by \citet{smithetal96} and \citet{Robertson2006}   at 25\degC\ with the current work.}
\label{fig:smithcomp}
\end{figure}

The focus of this work is the behaviour of neutral polymer solutions in the semidilute regime. DNA is a polyelectrolyte, so it is essential to ensure that sufficient salt is added to the DNA solutions such that all the charges are screened and they behave essentially like neutral synthetic polymer solutions. We have measured the hydrodynamic radius of two different linear DNA fragments across a range of salt concentrations (from 0.001 to 1 M) at 25\degC\, and the results are displayed \Figref{fig:salt1}. It is clear from the figure that complete charge screening occurs above 10 mM NaCl. This is in agreement with earlier dynamic light scattering studies on linear DNA \citep{Soda1984185,Langowski1987263,Liu20006001}. Since the solvent used here contains 0.5 M NaCl, the light scattering experiments of the current study are in a regime well above the threshold for observing charge screening effects. 

The effects of salt concentration on the persistence length of DNA has been studied earlier through a variety of techniques such as Flow Dichroism \citep{Rizzo19812143}, Flow Birefringence \citep{Cairney1982923}, Magnetic Birefringence \citep{Maret19832727} and force-extension experiments using optical tweezers \citep{Baumann19976185,Wenner20023160}. Data from a number of these studies has been collated in \Figref{fig:salt2}. Clearly, the persistence length in all the earlier studies appears to reach an approximately constant value of 45-50 nm for salt concentrations $\gtrsim 0.1$ M, suggesting that the charges have been fully screened in this concentration regime. The value of persistence length used in this study, $P=50$ nm (indicated by the dashed line in  \Figref{fig:salt2}), and the threshold concentration for charge screening obtained in this work, are consequently consistent with earlier observations in the high salt limit.

Figure~\ref{fig:smithcomp} compares present measurements of the dependence of hydrodynamic radius on molecular weight, with previous measurements~\citep{smithetal96,Robertson2006} at 25\degC. While \citet{smithetal96} used fragments and concatenates of $\lambda$ phage DNA to obtain molecules across the wide range of molecular weights that were studied, the measurements of \citet{Robertson2006} were carried out on molecules identical to those that have been used here. Both the earlier results were obtained by tracking fluorescently labeled linear DNA, in contrast to current measurements which were obtained by dynamic light scattering. The close agreement between results obtained by two entirely different techniques, across the entire range of molecular weights, establishes the reliability of the procedures adopted here. 

\begin{figure}[tbp]
\begin{center}
\epsfig{file=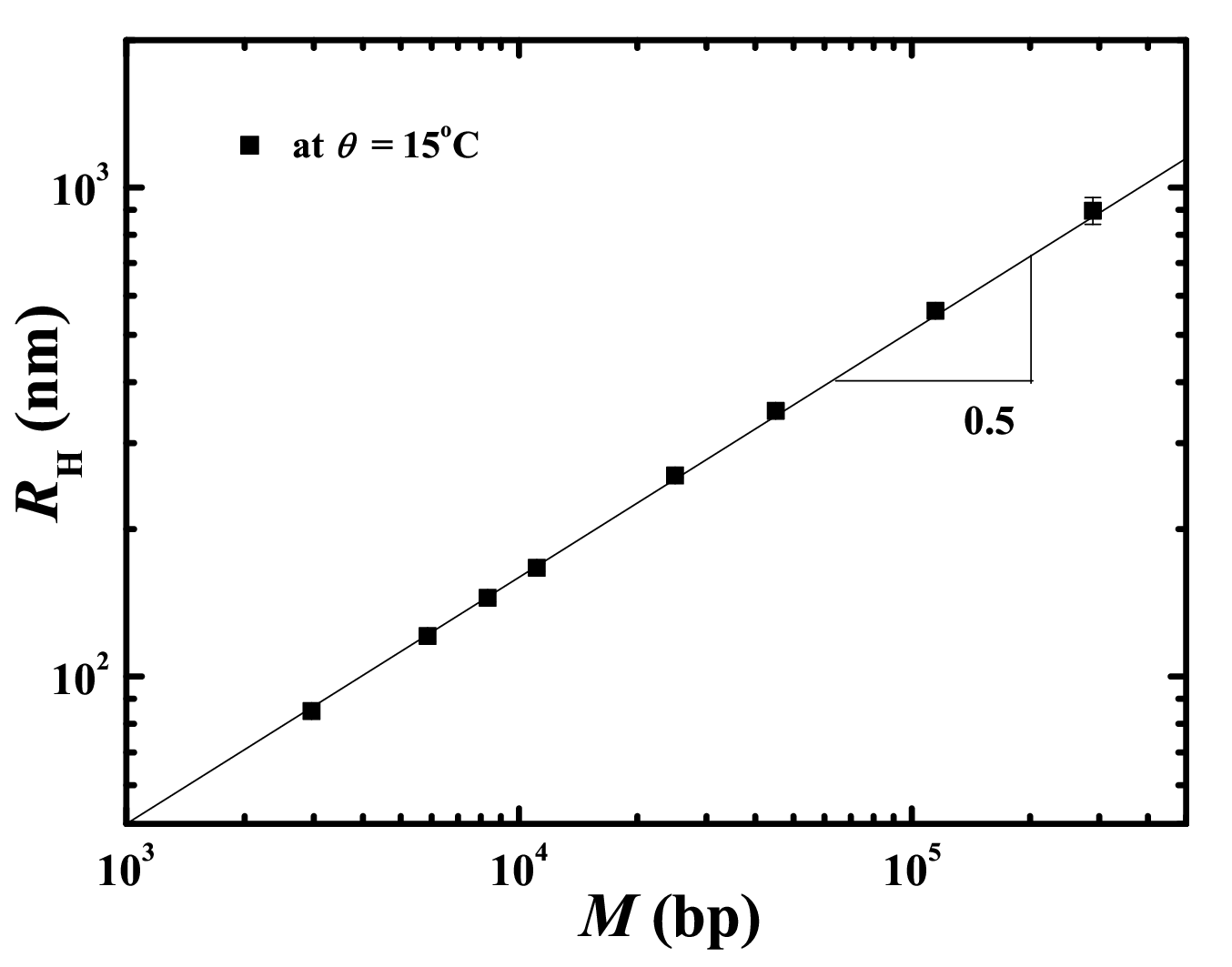,width=0.6\linewidth,clip=}
\end{center}
\caption{The variation of the hydrodynamic radius (\Rhth) with molecular weight (in bp) at $T = 15^{\circ}$C,  which is estimated to be close to the $\theta$-temperature.}
\label{fig:6new}
\end{figure}

\begin{table}[tbp]
\begin{center}
  \caption{Hydrodynamic Radius (\Rh) of linear DNA at different
temperatures. Each data point corresponds to the
intensity peaks from DLS measurements. The mean of 15 readings was
taken as final data point at each temperature for each DNA fragment. The values of \Rhth, with the $\theta$-temperature assumed to be $15^{\circ}$C,  are indicated in italics. 
}
\label{tab:RhSI}
\vskip10pt
\begin{tabular}{ccccc} \hline Sequence length & 2.9 kbp & 5.9 kbp &
8.3 kbp & 11.1 kbp \\ \hline Temperature & \Rh\ (in nm) & \Rh\ (in nm)
& \Rh\ (in nm) & \Rh\ (in nm)\\ \hline $5^{\circ}$C & 73$\pm$4 &
104$\pm$3 & 123$\pm$3 & 141$\pm$3\\ $10^{\circ}$C & 77$\pm$3 &
109$\pm$3 & 131$\pm$3 & 152$\pm$3\\ \emph{15\degC} & \emph{85$\pm$3} &
\emph{121$\pm$3} & \emph{145$\pm$3} & \emph{167$\pm$3}\\ $20^{\circ}$C & 87$\pm$3 &
124$\pm$3 & 148$\pm$3 & 173$\pm$4\\ $25^{\circ}$C & 90$\pm$3 &
131$\pm$5 & 155$\pm$2 & 183$\pm$6\\ $30^{\circ}$C & 96$\pm$2 &
136$\pm$4 & 162$\pm$3 & 189$\pm$3\\ $35^{\circ}$C & 101$\pm$4 &
145$\pm$7 & 174$\pm$3 & 203$\pm$5\\ \hline
\end{tabular}

\vspace{\baselineskip}
\begin{tabular}{ccccc} \hline Sequence length & 25 kbp & 45 kbp &
114.8 kbp & 289 kbp \\ \hline Temperature & \Rh\ (in nm) & \Rh\ (in
nm)& \Rh\ (in nm)& \Rh\ (in nm)\\ \hline $5^{\circ}$C & 203$\pm$4 &
258$\pm$5 & 385$\pm$13 & 540$\pm$35 \\ $10^{\circ}$C & 226$\pm$5 &
303$\pm$6 & 473$\pm$14 & 718$\pm$46 \\ \emph{15\degC} & \emph{258$\pm$3} &
\emph{349$\pm$4} & \emph{560$\pm$18} & \emph{897$\pm$57} \\ $20^{\circ}$C & 267$\pm$8 &
367$\pm$4 & 607$\pm$13 & 1025$\pm$39 \\ $25^{\circ}$C & 286$\pm$5 &
397$\pm$5 & 677$\pm$15 & 1201$\pm$49 \\ $30^{\circ}$C & 297$\pm$4 &
417$\pm$6 & 722$\pm$13 & 1300$\pm$38 \\ $35^{\circ}$C & 313$\pm$8 &
431$\pm$8 & 753$\pm$19 & 1363$\pm$57\\ \hline
\end{tabular}
\end{center}
\end{table}
Table \ref{tab:RhSI} is a compilation of all the results of measurements of  \Rh\ carried out here, across all molecular weights and temperatures. Since we have established that $T_{\theta} = 14.7\pm0.5\degC$, we expect the hydrodynamic radius to scale as $M^{0.5}$ at $T=15\degC$. \Figref{fig:6new} is a plot of \Rhth\ versus $M$, which clearly confirms that indeed ideal chain statistics are obeyed in the neighbourhood of the estimated $\theta$-temperature.

\begin{figure}[tbp]
\begin{center}
\includegraphics[width=0.75\textwidth]{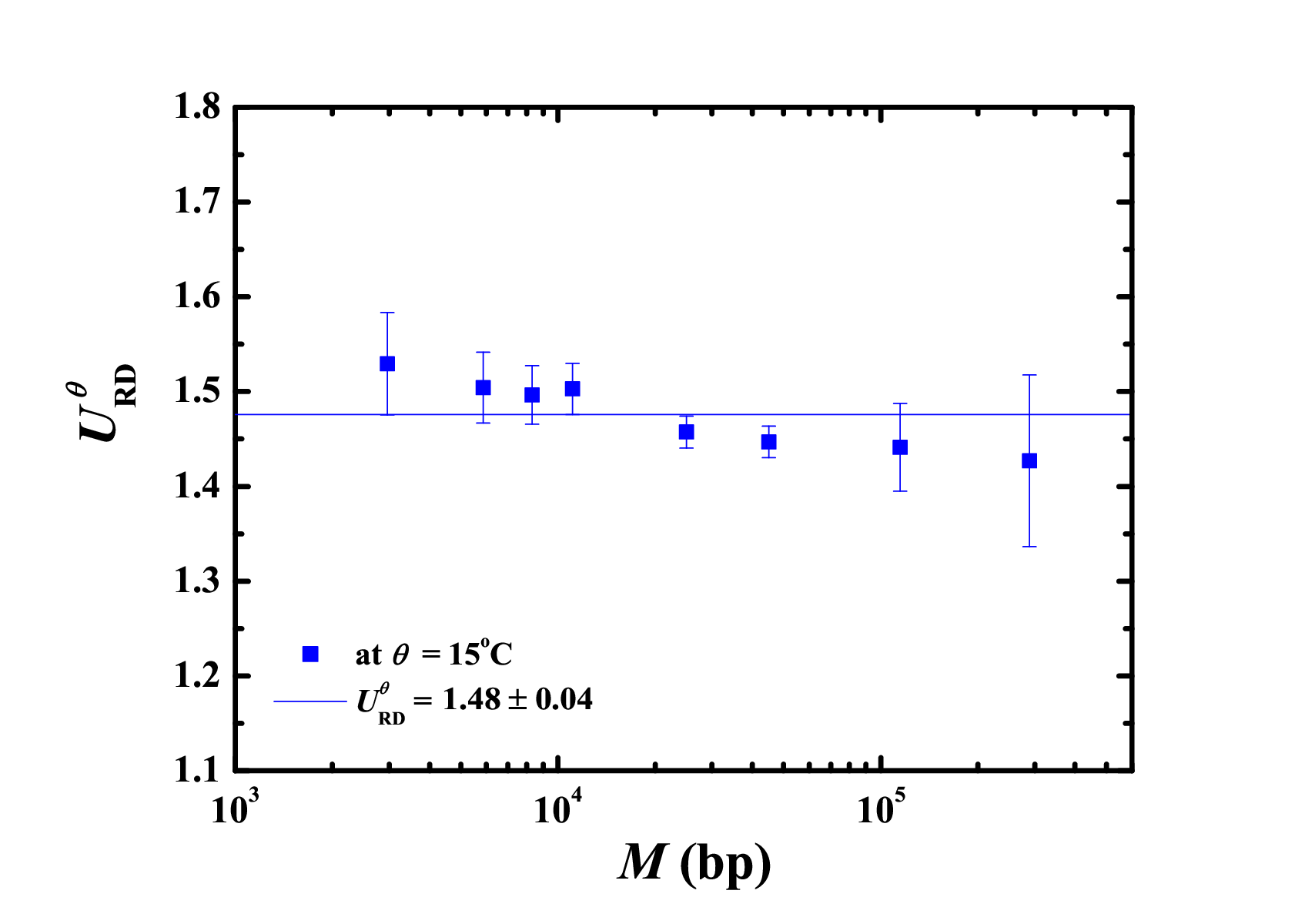}
\end{center}
\caption{The molecular weight independence of \urdth. The mean value is close to the Zimm model prediction in the long-chain limit, $\urdth \approx 1.47934$ \citep{Zimm1956,ottinger}.}
\label{fig:U_RD}
\end{figure}

Since both \Rgth\ and \Rhth\ scale with molecular weight as $M^{0.5}$ at the $\theta$-temperature, their ratio should be a constant. As is well known, experimental observations and theoretical predictions indicate that $\urdth = {\Rgth}/{\Rhth}$ is a chemistry independent universal constant (for a recent compilation of values see Table~I in~\cite{Kroger20004767}). Zimm theory predicts a universal value $\urdth \approx$ 1.47934~\citep{Zimm1956,ottinger}. Since we have estimated \Rgth\ by assuming Gaussian chain statistics at the $\theta$ temperature, and have measured \Rhth, we can calculate \urdth\ for all the molecular weights used in this work. The expected molecular weight independence of \urdth\ is displayed in \Figref{fig:U_RD}. The mean value of \urdth\ is also seen to be close to the value predicted by Zimm. This confirms that both the scaling with molecular weight, and the absolute values of \Rhth, across the entire range of DNA molecular weights, are accurately captured by the dynamic light scattering experiments. 

\begin{figure}[tbp]  \centering
\includegraphics[width=0.61\textwidth]{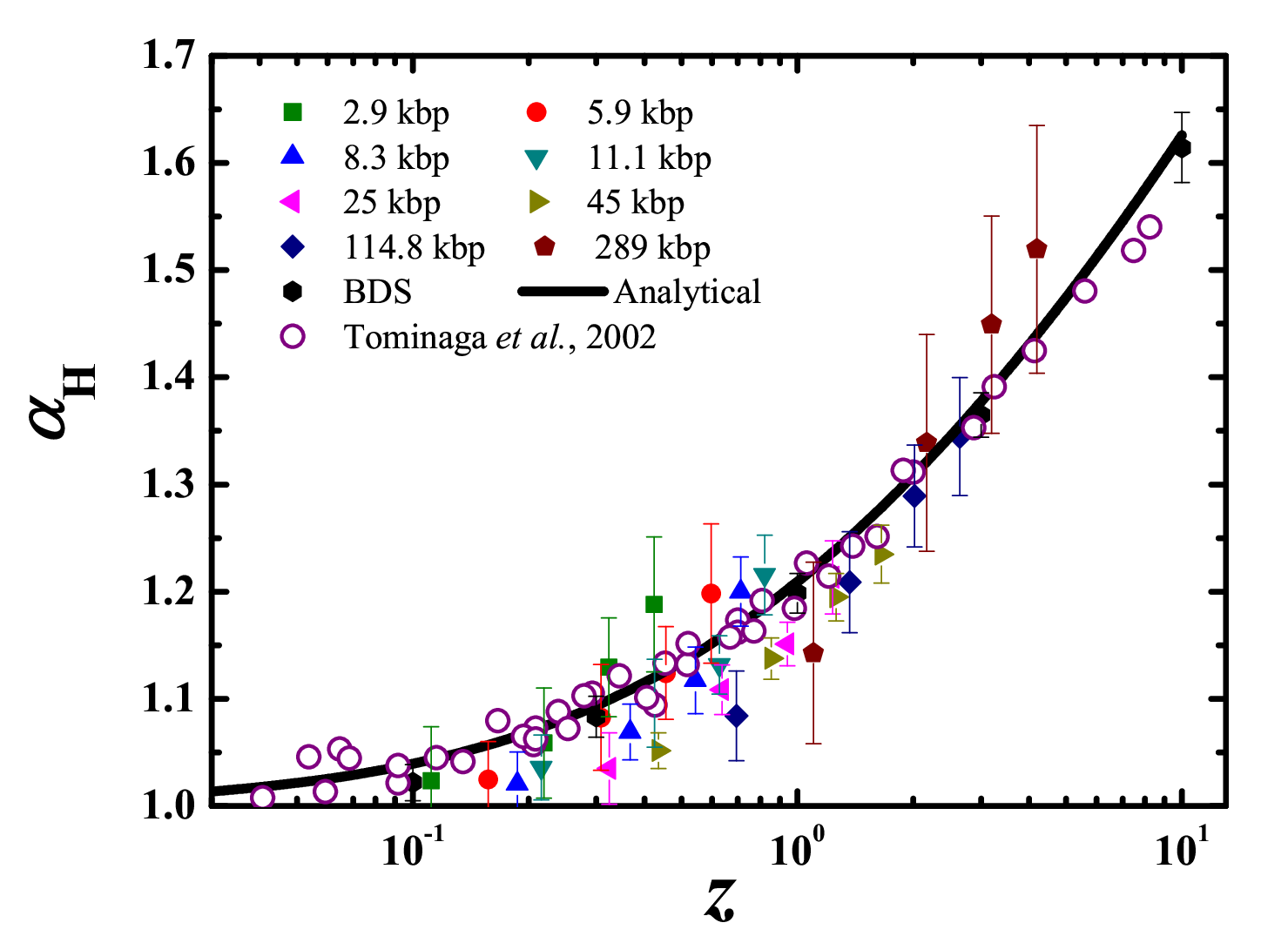}
\caption{Swelling of the hydrodynamic radius. The filled coloured
symbols represent experimental data for DNA. BDS refers to the predictions of Brownian dynamics
simulations~\citep{SunRav06-epl}, with the curve representing the function $f_\text{H}(z)$, with constants $a = 9.528$, $b = 19.48$, $c = 14.92$, and $m = 0.0999$. Empty circles represent several experimental data on synthetic polymers collated in \citet{Tominaga20021381}.}
\label{fig:alphaHvsz}
\end{figure}

The swelling  $\alpha_{\mathrm{H}}$ for any combination of $M$ and $T$ can be calculated from the values reported in Table \ref{tab:RhSI}, and plotted as a function of the scaling variable $z = k\left(1- {T_{\theta}}/{T} \right) \!\sqrt{M}$, once a choice has been made for the value of the constant $k$. As mentioned earlier, $k$ can be determined by comparison of experimental measurements with the results of Brownian dynamics simulations. We refer the interested reader to the relevant literature~\citep{Domb1976179,Barrett1991,Yama01,Schafer99,Kumar20037842,SunRav06-epl} for a discussion of how the solvent quality parameter $z$ enters the structure of analytical theories and Brownian dynamics simulations. It suffices here to note that the theoretically predicted swelling of the hydrodynamic radius can be represented by the functional form $\ah = f_\text{H} (z)$, where, $f_\text{H}(z) = (1 + a\, z + b\, z^{2} + c \, z^{3})^{m/2}$, with the values of the constants $a$, $b$, $c$, $m$, etc., dependent on the particular context. The values of the various constants that fit the results of Brownian dynamics simulations, are reported in the caption to Fig.~\ref{fig:alphaHvsz}. We find the constant $k$ for DNA solutions by adopting the following procedure. 

Consider $\ah^\text{expt}$ to be the experimental value of swelling at a particular value of temperature $T$ and molecular weight $M$. It is then possible to find the Brownian dynamics value of $z$ that would give rise to the same value of swelling from the expression $z = f^{-1}_\text{H} (\ah^\text{expt})$, where $f^{-1}_\text{H}$ is the inverse of the function $f_\text{H}$. Since $z = k \, \hat \tau \, \sqrt{M}$, where $\hat \tau=  \left(1- \dfrac{T_{\theta}}{T} \right)$, it follows that a plot of $f^{-1}_\text{H}(\ah^\text{expt})/\sqrt{M}$ versus $\hat \tau$, obtained by  using a number of values of $\ah^\text{expt}$ at various values of $T$ and $M$, would be a straight line with slope $k$. Once the constant $k$ is determined, both experimental measurements of swelling and results of Brownian dynamics simulations can be represented on the same plot. Assuming that the $\theta$-temperature is 15\degC\ for the solvent used in this study, we have determined the value of $k$ by following this procedure (see supplementary material for greater detail). It follows that for any given molecular weight and temperature, the solvent quality $z$ for the DNA solution can be determined. Typical values of $z$, at various $M$ and $T$, obtained by this procedure are reported in Table~\ref{tab:zc}.

\begin{figure}[t]
\begin{center}
\epsfig{file=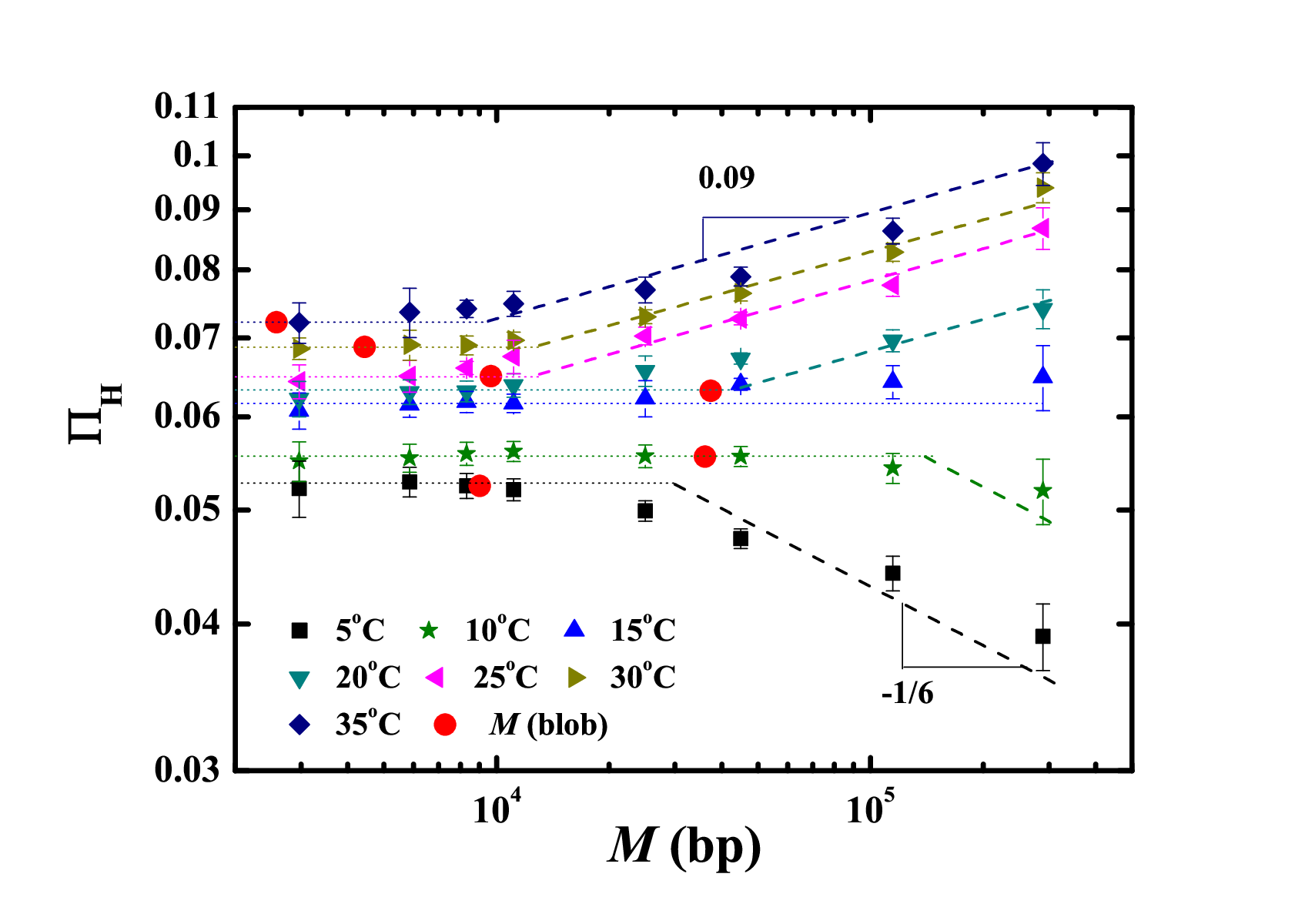,width=0.7\linewidth,clip=}
\end{center}
\caption{Different scaling regimes for the scaled variable \pih\ as a function of molecular weight $M$. The filled red circles correspond to the molecular weight $M_\text{blob}$ of the chain segment within a thermal blob.}
\label{fig:logpihblob}
\end{figure}

The solvent quality crossover of  \ah\ for DNA, determined from the current measurements, is shown in Fig~\ref{fig:alphaHvsz}, along with the predictions of Brownian dynamics simulations.  Experimental data of \citet{Tominaga20021381}, which are considered to be highly accurate measurements of synthetic polymer swelling, are also plotted in the same figure. It is evident from the figure that, just as in the case of synthetic polymer solutions, irrespective of solvent chemistry, the swelling of DNA is universal in the  crossover region between $\theta$ to good solvents. 

Having estimated the value of $z$ for any values of $M$ and $T$, it follows that other universal properties predicted by simulations or theory, at any particular value of $z$, can be compared with experimental results for DNA,  at the same value of $z$.

\section{\label{sec:blob} Thermal blobs and measurements in poor solvents}

The focus of the experimental measurements in the dilute limit reported in Appendices~\ref{sec:sls} and \ref{sec:dls} is twofold: (i) determining the $\theta$-temperature, and (ii) describing the $\theta$ to good solvent crossover behaviour of a solution of double-stranded DNA. The analysis of properties under poor solvent conditions has been carried out essentially  only in order to locate the $\theta$-temperature. As is well known, the experimental observation of single chains in poor solvents is extremely difficult because of the problem of aggregation due to interchain attraction. Nevertheless, in this section we show that a careful analysis of the dynamic light scattering data, in the light of the blob picture, enables us to discuss the reliability of the measurements that have been carried out here under poor solvent conditions.

According to the blob picture of dilute polymer solutions, a polymer chain in a good or poor solvent can be considered to be a sequence of thermal blobs, where the thermal blob denotes the length scale at which excluded volume interactions become of order $k_\text{B} T$~\citep{RubCol03}. Under good solvent conditions, the blobs obeying self-avoiding-walk statistics, while they are space filling in poor solvents. As a result, the mean size $R$ of a polymer chain (assumed here to be the magnitude of the end-to-end vector) is given by~\citep{RubCol03},
\begin{equation}
R =   R_\text{blob} (T) \left(\frac{N_{k}}{N_\text{blob} (T) } \right)^{\nu}
\label{eq:chainsize}
\end{equation}
where, $N_{k}$ is the number of Kuhn-steps in a chain, $N_\text{blob}$ is the number of Kuhn-steps in a thermal blob, and $R_\text{blob}$ is the mean size of a thermal blob. The Flory exponent $\nu$ is $\approx 0.59$ in a good solvent, and $1/3$ in a poor solvent. The size of the thermal blob is a function of temperature. For instance, under athermal solvent conditions, the entire chain obeys self avoiding walk statistics, so the blob size is equal to the size of a single Kuhn-step. On the other hand, for temperatures approaching the $\theta$-temperature, the blob size grows to engulf the entire chain. 

It is convenient to define the following dimensionless scaling variable:
\begin{equation}
  \label{eq:pih} \pih \equiv \frac{\Rh}{ a \, \sqrt{M}}
\end{equation}
where, $a$ is a constant with dimensions of length, which we have set equal to 1 nm. In general,
\pih\ should increase with molecular weight for good solvents, remain constant for theta
solvents, and decrease for poor solvents. However, Eq.~(\ref{eq:chainsize}) suggests that on length scales smaller than the blob length scale \pih\ must remain constant, while on length scales large compared to the blob length scale, \pih\ must scale as $M^{0.09}$ in good solvents, and $M^{-1/6}$ in poor solvents. Figure~\ref{fig:logpihblob} is a plot of $\log \pih$ versus $\log M$, obtained from the measurements carried out in this study, in the light of these arguments. It is clear that after an initial regime of constant values, there is a crossover to the expected scaling laws in both the good and poor solvent regimes. The crossover from one scaling regime to the next begins approximately at the blob length scale, an estimate of which can be made as follows.

\begin{table}[t]
\caption{Equations for the dimensionless excluded volume parameter $v_{0}/b_{k}^{3}$, and the molecular weight of the chain segment within a thermal blob $M_\text{blob}$, in good and poor solvents. Here, $m_{k}$ is the molar mass of a Kuhn-step, and $U_{R}$ and $U_{RD}$ are universal amplitude ratios, such that $R = U_{R} \, \Rg$, and $\Rg = U_{RD} \, \Rh$. In all the calculations here, we assume $U_{R} = \sqrt{6}$, and $U_{RD} = 1.46$.} 
\label{tab:eqns}
\vskip10pt
\begin{tabular}{| c | c | c |}
\hline
solvent quality      & good
            & poor
            \\
\hline
\hline
$\dfrac{v_{0}}{b_{k}^{3}}$   &  $\left[ \dfrac{a \, \pih  \, (U_{R} U_{RD}) \, m_{k}^{\nu} }{b_{k}}  \right]^{\tfrac{1}{2 \nu -1}} \dfrac{1}{M^{\tfrac{1}{2}} \left(1- \dfrac{T_\theta}{T} \right)}$
             &$\left[ \dfrac{a \, \pih  \, (U_{R}U_{RD})  }{b_{k}}  \right]^{-3} \dfrac{1}{m_{k} \,  M^{\tfrac{1}{2}} \left(1 - \dfrac{T}{T_\theta}\right)}$
\\
(for $M > M_\text{blob}$) &&\\
\hline
$M_\text{blob} (T)  $     & $\dfrac{m_{k} \, b_{k}^{6}}{v^{2}_{0} \left(1- \dfrac{T_\theta}{T} \right)^{2}} $
            & $\dfrac{m_{k} \, b_{k}^{6}}{v^{2}_{0}  \left(1 - \dfrac{T}{T_\theta} \right)^{2}}     $      \\
\hline
\end{tabular}
\end{table}

\begin{figure}[t]
\begin{center}
\epsfig{file=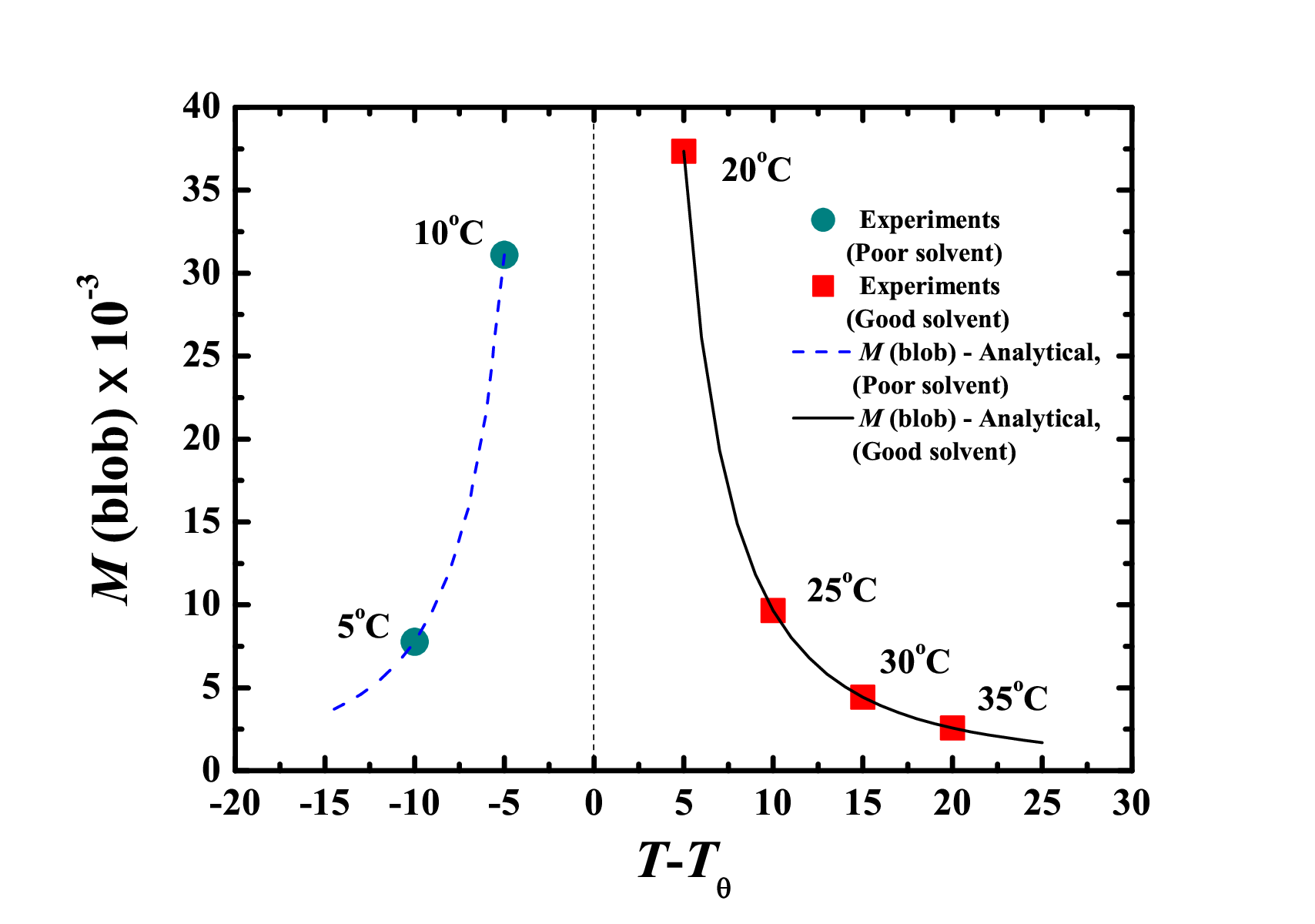,width=0.7\linewidth,clip=}
\end{center}
\caption{Variation of the molecular weight of the chain segment within a thermal blob  with respect to temperature, on either side of the $\theta$-temperature (see Table~\ref{tab:eqns} for the equations governing $M_\text{blob}$). The symbols denote values at temperatures at which experimental measurements have been made.}
\label{fig:Mblob}
\end{figure}

The requirement that the energy of excluded volume interactions within a thermal blob are of order $k_\text{B} T$ leads to the following expressions for  $N_\text{blob}$ and $R_\text{blob}$~\citep{RubCol03},
\begin{align}
  \label{eq:Nblob} N_\text{blob} (T) = & \frac{b_{k}^{6}}{ {v(T)}^{2}} \\
  \label{eq:Rblob} R_\text{blob} (T) =  &   \frac{b_{k}^{4}}{ \left\vert v(T) \right\vert }
\end{align}
where, $b_{k}$ is the length of a Kuhn-step, and $v (T)$ is the excluded volume at temperature $T$.  The excluded volume can be shown to be related to the temperature through the relation,
\begin{equation}
v (T) = \begin{cases}
     v_{0} \left(1- \dfrac{T_\theta}{T} \right) & \text{for good solvents}, \\
     & \\
    -  v_{0}  \left( 1 - \dfrac{T}{T_\theta}  \right) & \text{for poor solvents}.
\end{cases}
\label{eq:v0}
\end{equation}
where, $v_{0}$ is a chemistry dependent constant. These expressions are consistent with the expectation that $v \to v_{0}$ in an athermal solvent ($T \to \infty$), and   $v \to - v_{0}$ in a non-solvent ($T \to 0$)~\citep{RubCol03}. Since measurements of  the mean size (via \Rh) have been carried out here at various temperatures, and we have estimated both $T_\theta$  and $b_{k}$, it is possible to calculate $v_{0}$ using Eqs.~(\ref{eq:chainsize}) to (\ref{eq:Rblob}). As a result the size of a thermal blob as a function of temperature can also be estimated.

The equations that govern the dimensionless excluded volume parameter $v_{0}/b_{k}^{3}$ and the molecular weight $M_\text{blob}$ of a chain segment within a thermal blob, in good and poor solvents, are tabulated in Table~\ref{tab:eqns}, when the hydrodynamic radius \Rh\ is used as a measure of chain size. Here, $m_{k}$ is the molar mass of a Kuhn-step, and the universal amplitude ratio $U_{R}$ has been used to relate the magnitude of the end-to-end vector $R$ to \Rg\ ($R = U_{R} \, \Rg$), while the universal ratio $U_{RD}$  relates \Rg\ to \Rh\ ($\Rg = U_{RD} \, \Rh$). The values of these ratios are known analytically for the case of Gaussian chains and Zimm hydrodynamics under $\theta$-conditions~\citep{DoiEd86}, and numerically in the case of good solvents~\citep{Kumar20037842}, and when fluctuating hydrodynamic interactions are taken into account~\citep{SunRav06-epl}.

\begin{figure}[!ht]
\centering 
\subfloat[][]{ 
    \includegraphics[width=0.65\textwidth]{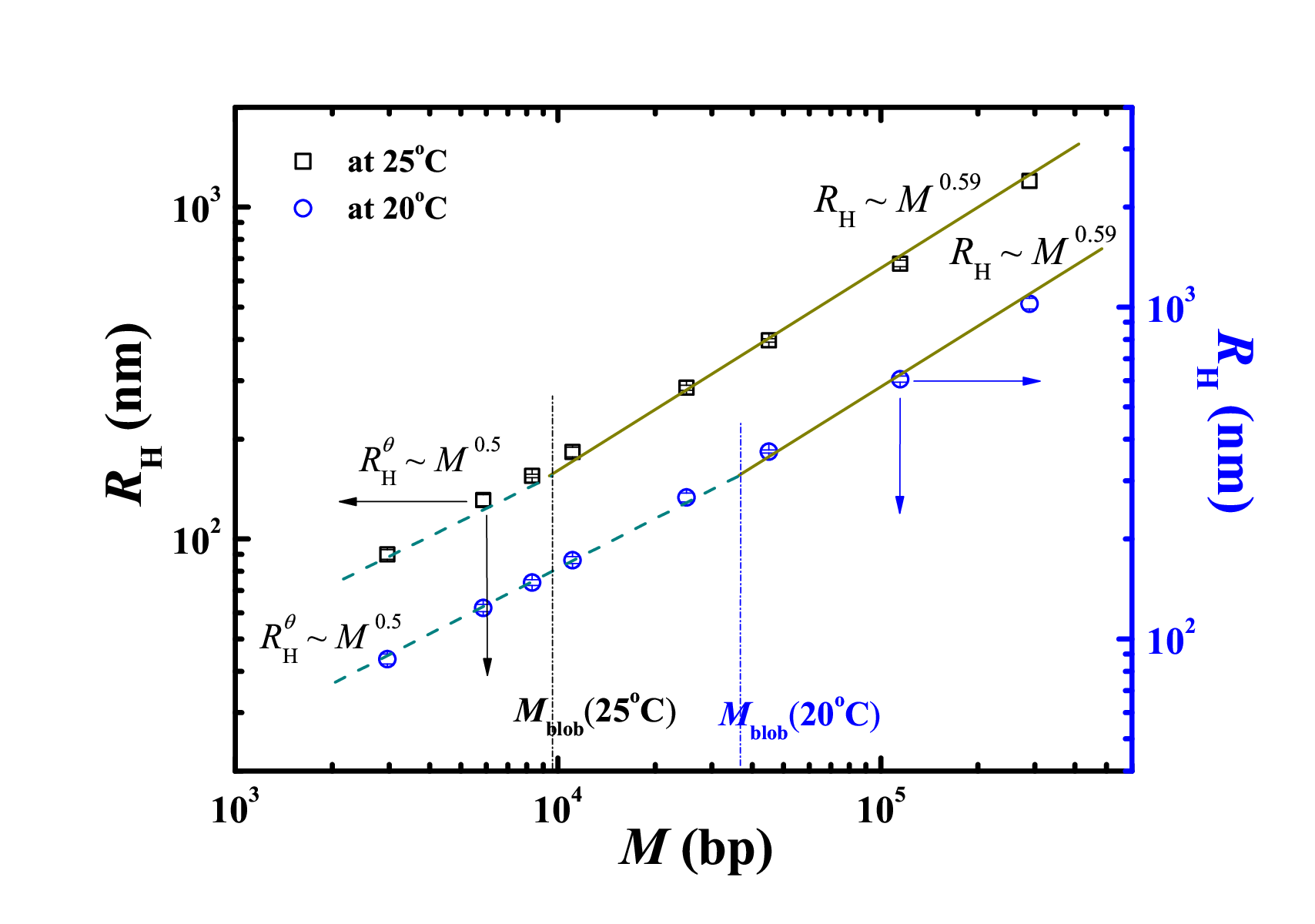}
  \label{figgood:sub:a}}    \\
\subfloat[][]{ 
    \includegraphics[width=0.65\textwidth]{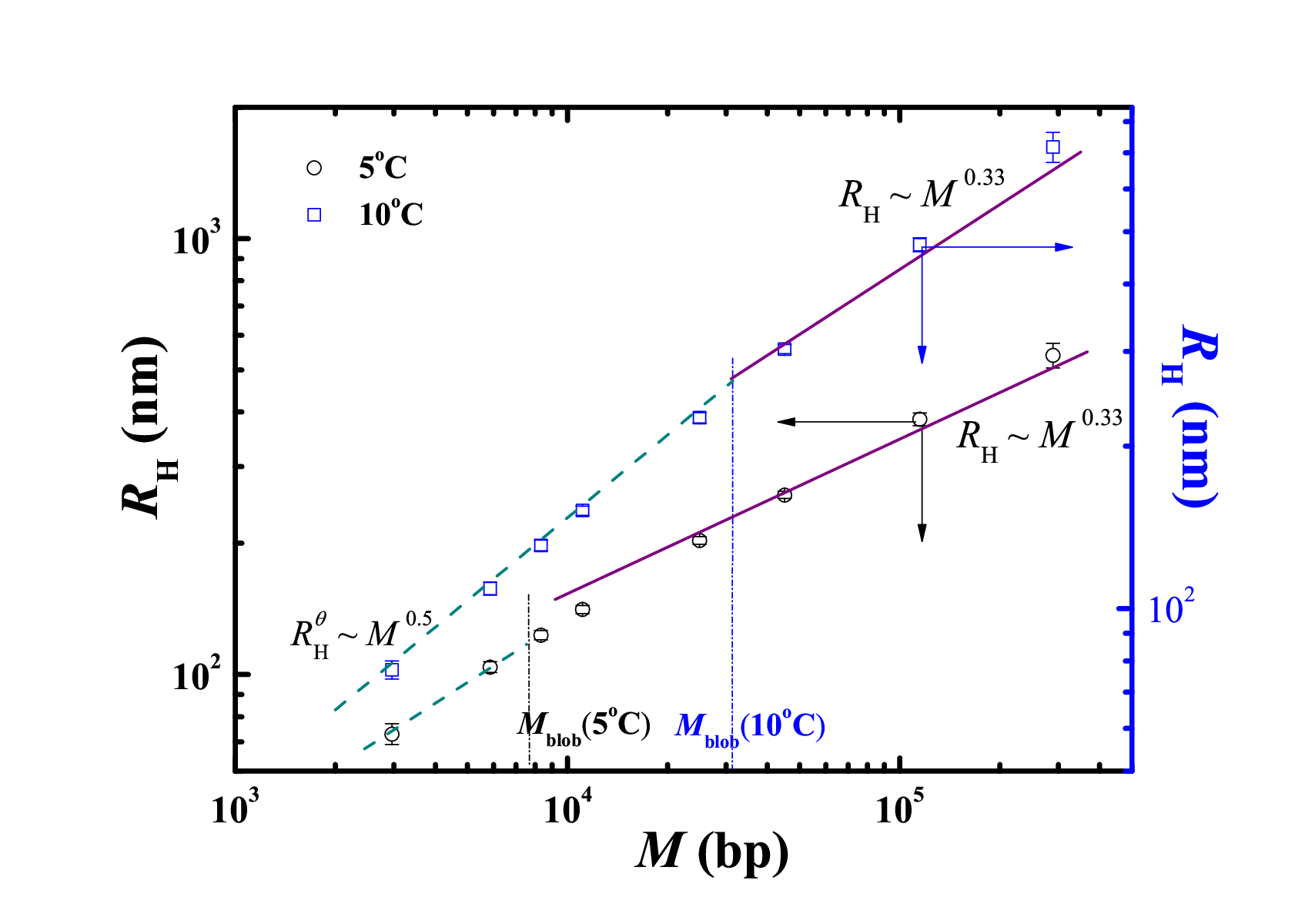}
   \label{figpoor:sub:b}}
%\begin{spacing}{1.5}
\caption{The variation of hydrodynamic radius (\Rh) with molecular weight (in bp) in (a) good solvents at $20^{\circ}$C and $25^{\circ}$C, and (b) poor solvents at $5^{\circ}$C and $10^{\circ}$C. The scaling of \Rh\ with $M$ appears to obey Gaussian statistics within the thermal blob, self-avoiding walk statistics for $M > M_\text{blob}$ in good solvents, and collapsed globule statistics for $M > M_\text{blob}$ in poor solvents.}
\label{fig:blobrh}
\end{figure}

Using the known values of $a, \pih, b_{k}, m_{k}, U_{RD}, U_{R}$ in the appropriate equations in Table~\ref{tab:eqns}, we find that for sufficiently high molecular weights, $v_{0}/b_{k}^{3} \approx 5.4 \pm 0.2$ in \textit{both} good and poor solvents. This is significant since an inaccurate measurement of mean size in a poor solvent (as a consequence of, for instance, chain aggregation), would result in different values of $v_{0}/b_{k}^{3}$ in good and poor solvents. Further evidence regarding the reliability of poor solvent measurements can be obtained by calculating $M_\text{blob} (T)$ in good and poor solvents.

Figure~\ref{fig:Mblob} displays the variation of $M_\text{blob}$ with respect to the temperature difference $T- T_{\theta}$, calculated using the equations given in Table~\ref{tab:eqns}. The figure graphically demonstrates the temperature dependence of the blob size, and confirms that essentially the blob size is the same in either a good or poor solvent when the temperature is equidistant from the $\theta$-temperature. The symbols in Fig.~\ref{fig:Mblob} denote values of $M_\text{blob}$, evaluated at the temperatures at which experimental measurements have been made. These values have been represented by the filled red circles in Fig.~\ref{fig:logpihblob}. As can be seen from Fig.~\ref{fig:logpihblob}, the magnitude of $M_\text{blob}$ is roughly consistent with the location of the crossover from the scaling regime within a blob, to the scaling regime that holds at length scales larger than the blob, in both good and poor solvents. The two scaling regimes, in good and poor solvents,  are illustrated explicitly in Figs.~\ref{fig:blobrh}.

\begin{figure}[t]
\begin{center}
\epsfig{file=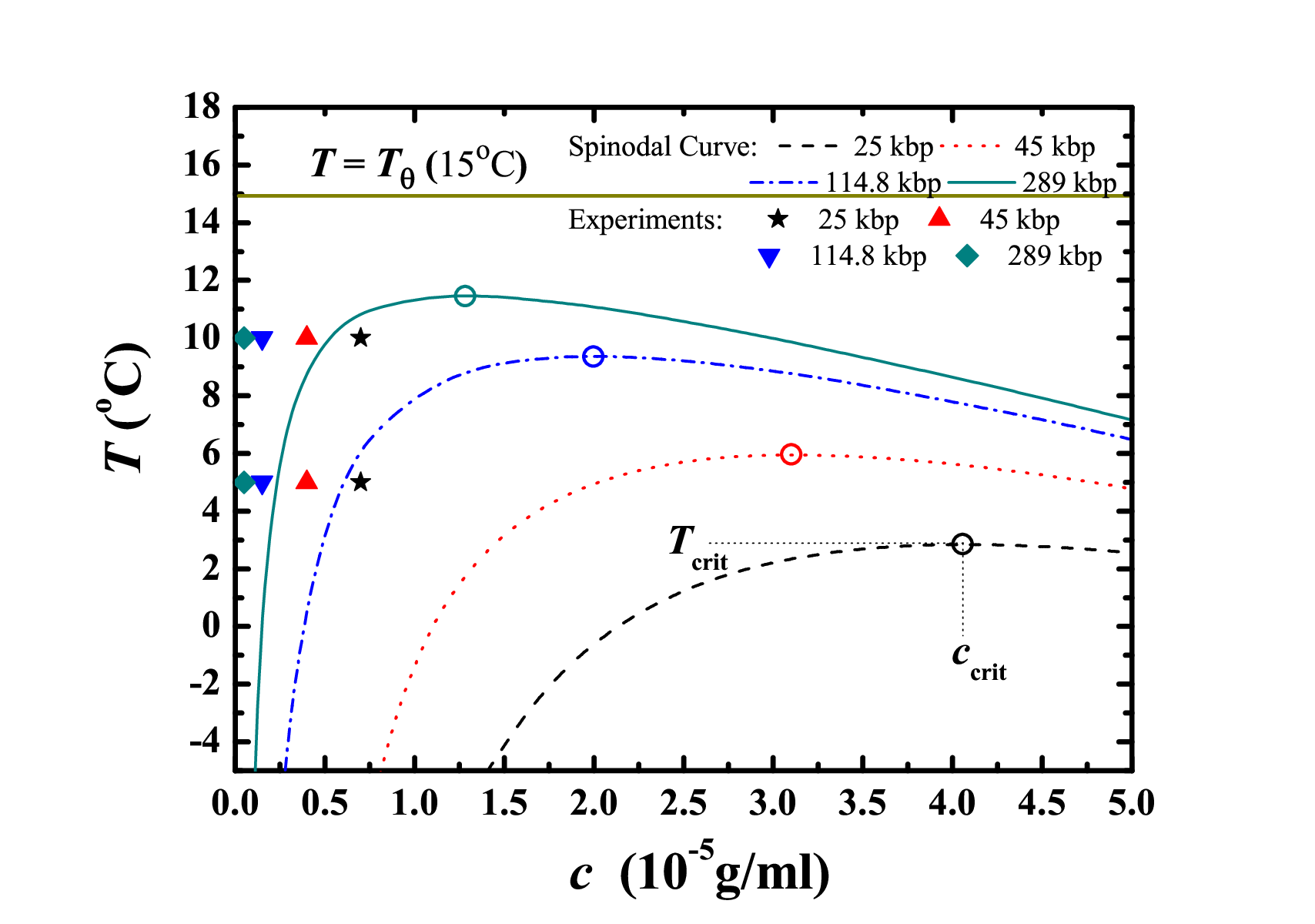,width=0.7\linewidth,clip=}
\end{center}
\caption{Spinodal curves and critical temperatures and concentrations (filled circles) predicted by Flory-Huggins mean-field theory for a range of molecular weights. Values of concentrations and temperatures at which the poor solvent experiments have been conducted are also indicated.}
\label{fig:FH}
\end{figure}

The possibility of phase separation under poor solvent conditions, as polymer-solvent interactions become less favourable, is the primary reason for the difficulty of accurately measuring the size scaling of single chains. An approximate estimate of the thermodynamic driving force for phase separation can be obtained with the help of Flory-Huggins mean field theory. Since the Flory-Huggins $\chi$ parameter is related to the excluded volume parameter through the relation~\citep{RubCol03} $\chi = \dfrac{1}{2}\left[1- \dfrac{v (T)}{b_{k}^{3}} \right]$, and we have estimated the value of $v (T)$ in both solvents, the phase diagram predicted by Flory-Huggins theory for dilute DNA solutions considered here can be obtained. It is appropriate to note that we are not interested in accurately mapping out the phase diagram for DNA solutions with the help of Flory-Huggins theory. This has already been studied in great detail, using sophisticated versions of mean-field theory, starting with the pioneering work of \cite{Post1982}, and the problem of DNA condensation is an active field of research~\citep{Yoshikawa1996,yoshikawa2011,Teif2011}. Our primary interest is to obtain an approximate estimate of the location of the current experimental measurements relative to the unstable two-phase region (whose boundary is determined by the spinodal curve), since phase separation can occur spontaneously within this region. Figure~\ref{fig:FH} displays the spinodal curves for the 25 to 289 kbp molecular weight samples, predicted by Flory-Huggins theory, using the parameters for the current measurements. Details of how these curves can be obtained are given, for instance, in \cite{RubCol03}. Also indicated on each curve are the critical concentration and temperature. It is clear by considering the location of the symbols denoting the concentration-temperature coordinates of the poor solvent experiments, that for each molecular weight, they are located outside the unstable two-phase region, lending some justification to the reliability of the present poor solvent measurements. It is appropriate to note here that mean-field theories do not accurately predict the shape of the binodal curve, and in general concentration fluctuations tend to make the curve wider close to the critical point~\citep{RubCol03}. Interestingly, even for the 289 kbp sample, that has a very large molecular weight ($\approx 1.9 \times 10^{8}$ Dalton), there is still a considerable gap between the critical and $\theta$-temperatures ($\approx 4^{\circ}$C). The reason for this is because the stiffness of double-stranded DNA leads to a relatively small number of Kuhn-steps (983) even at this large value of molecular weight, and the value of the critical temperature predicted by Flory-Huggins theory depends on the number of Kuhn-steps in a chain rather than the molecular weight.

\bibliography{jorrefs}
\noindent
\bibliographystyle{JORnat}

\end{document}

\title{Supplementary material for: \\
Universal solvent quality crossover of the zero shear rate viscosity of semidilute DNA solutions }
\date{\today}
\author{Sharadwata Pan}
\affiliation{IITB-Monash Research Academy, Indian Institute of Technology Bombay, Powai, Mumbai - 400076, India}
\affiliation{Department of Chemical Engineering, Indian Institute of Technology Bombay, Powai, Mumbai - 400076, India}
\affiliation{Department of Chemical Engineering, Monash University, Melbourne, VIC 3800, Australia}
\author{Duc At Nguyen}
\affiliation{Department of Chemical Engineering, Monash University,
Melbourne, VIC 3800, Australia}
\author{T. Sridhar}
\affiliation{Department of Chemical Engineering, Monash University, Melbourne, VIC 3800, Australia}
\affiliation{IITB-Monash Research Academy, Indian Institute of Technology Bombay, Powai, Mumbai - 400076, India}
\author{P. Sunthar}
\affiliation{Department of Chemical Engineering, Indian Institute of Technology Bombay, Powai, Mumbai - 400076, India}
\affiliation{IITB-Monash Research Academy, Indian Institute of Technology Bombay, Powai, Mumbai - 400076, India}
\author{J. Ravi Prakash}
\email[Corresponding author: ]{ravi.jagadeeshan@monash.edu}
%\homepage[Visit:]{http://users.monash.edu.au/~rprakash/}
\affiliation{Department of Chemical Engineering, Monash University, Melbourne, VIC 3800, Australia}
\affiliation{IITB-Monash Research Academy, Indian Institute of Technology Bombay, Powai, Mumbai - 400076, India}

\maketitle

\section{Details of strains, working conditions and procedures for
  preparing linear DNA fragments.}

Recently, a range of special DNA constructs, from $3-300$ kbp, have
been genetically engineered into bacterial strains of \emph{E. coli}
\citep{Laib20064115}, which can be selectively extracted for
rheological studies. Primarily they fall into three categories:
plasmids, fosmids and Bacterial Artificial Chromosomes
(BAC). Altogether six samples (two plasmids, two fosmids and two
BACs), which were originally prepared elsewhere \citep{Laib20064115}
were procured from Dr. Brad Olsen, California Institute of Technology,
USA. Throughout the work, the nomenclature of all the three types of
DNA samples has been used as in the originally published
work \citep{Laib20064115}. In addition, two special bacterial strains
containing the plasmids: pBSKS (2.9 kbp) and pHCMC05 (8.3 kbp) were
provided by Dr. S. Noronha, Dept. of Chemical Engineering, IIT Bombay, India.
The details about size, growth conditions of bacteria and
single cutters of the DNA samples are given in
\Tabref{tab:dna}. After procurement of samples (in the form of agar
stab cultures of \emph{E. coli}), glycerol freeze stocks were made
using 50\% glycerol and stored at -80\degC. The cultures can be stored
in this way for several years and can be used at any time to produce
DNA samples \citep{Laib20064115}.

\begin{table}[tbp]
\begin{center}
  \caption[DNA Fragments] {DNA Fragments. Here `LB' stands for Luria
    Bertini broth, `Ant$^{R}$' refers to Antibiotic resistance, `Amp.'
    refers to Ampicillin, `CAM' refers to Chloramphenicol, `Kan'
    refers to Kanamycin, all the cultures were incubated overnight at
    $37^{\circ}$C with vigorous shaking (200--250 rpm). L-arabinose
    (inducer) was used at a concentration of 0.01g per 100 ml (stock
    concentration: 5g in 100ml). Stock concentrations for preparations
    of Ampicillin, Chloramphenicol and Kanamycin were 100 mg/ml, 25
    mg/ml and 100 mg/ml respectively. The working concentrations for
    Amp., CAM and Kan are 100 $\mu$g/ml, 12.5$\mu$g/ml and
    100$\mu$g/ml respectively. Growth conditions for all the plasmids
    are same (LB + Amp.) except pHCMC05 (LB + Amp. + CAM). For both
    the fosmids, growth conditions are identical (LB + CAM +
    L-arabinose). For both the BACs, growth conditions are same (LB +
    CAM + Kan + L-arabinose).}
\label{tab:dna}
\vskip10pt
\begin{tabular}{ccccc}
 \hline
      Type & Name & Size (kb) /(Notation)& Ant$^{R}$ & 1 Cutter \\ \hline
      Plasmid
            & pBSKS(+)   & 2.9 / F2.9 &  Amp. & BamHI \\
            & pYES2   & 5.9 / F5.9 & Amp. & BamHI  \\
            & pHCMC05   & 8.3 / F8.3 &  Amp. + CAM & BamHI  \\
            & pPIC9K$<$TRL5$>$   & 11.1 / F11.1 &  Amp. & BamHI\\
    \hline
    Fosmid & pCC1FOS-25   & 25 / F25 &  CAM & ApaI \\
           & pCC1FOS-45   & 45 / F45 &  CAM & ApaI \\
    \hline
    BAC & CTD-2342K16   & 114.8 / F114.8 &  CAM + Kan & MluI \\
        & CTD-2657L24   & 289 / F289 & CAM + Kan & MluI \\
   \hline
\end{tabular}
\end{center}
\end{table}

Standard procedures \citep{Laib20064115, Sambrook2001} involving
alkaline lysis (mediated by NaOH) were adopted for extraction,
linearization and purification of plasmids, fosmids and BAC from the
cultures. For high copy number plasmids, no inducer was added. For low
(fosmids) and very low (BACs) copy number samples, L-arabinose was
added as inducer. From each freeze stock, 15$\mu$l of ice was scrapped
and transferred to 40 ml LB medium with proper antibiotic (as
mentioned in \Tabref{tab:dna}) and incubated overnight ($16-18$ hours)
at 37\degC\ with vigorous shaking ($200-250$ rpm). The overnight grown
culture was poured into microcentrifuge tubes and cells were harvested
by centrifugation. The bacterial pellet (obtained above) was
resuspended in 100$\mu$l of ice-cold Solution I (4\degC)
 followed by 200$\mu$l of freshly prepared
Solution II  and 150$\mu$l of ice-cold Solution
III \citep{Sambrook2001}. The tubes were stored on ice for 3--5
minutes and centrifuged. The supernatant was transferred to a fresh
tube. The precipitate (containing mainly the cell debris and genomic
DNA) was discarded. RNase was added (at 10 $\mu$g/ml) to the tube and
incubated at 37\degC\ for 20 minutes. Equal volume of
Phenol-Chloroform-Isoamyl Alcohol (25:24:1) mixture was added and
mixed well by vortexing. After centrifugation, supernatant was
transferred to a fresh tube. Equal volume of Chloroform was added and
centrifuged. The supernatant was transferred to a fresh tube. Two
volumes of chilled 100\% ethanol (at 4\degC) was added at room
temperature kept for 7--8 hours at -20\degC. The tube was then
centrifuged and the supernatant removed by gentle aspiration. The tube
was kept in an inverted position in a paper towel to allow all of the
fluid to drain away. Following this, 1 ml of 70\% ethanol was added to
the tube and centrifuged. When all of the ethanol was evaporated, the
resulting DNA pellet was dissolved in 50$\mu$l of Milli-Q grade water
and stored at -20\degC.

To linearize the extracted DNA fragments, 39$\mu$l of water was added
to a 1.7 ml microcentrifuge tube, followed by 10$\mu$l of
corresponding 10X Assay Buffer (working concentration is 1X) and
50$\mu$l of DNA solution (purified DNA stored at 4\degC). 1 $\mu$l of
appropriate enzyme was added. A thumb rule is 0.5 - 1 U enzyme for 1
mg DNA \citep{Sambrook2001}. The samples were mixed well with
micropippette (wide bore tips) for several times. The reaction mix
(100$\mu$l) was incubated at 37\degC\ for three hours. After
restriction digestion / linearization, it is necessary to remove the
enzymes / other reagents present in the reaction mix so that they do
not interfere with the downstream application/s like light scattering
studies, rheometry etc. For this normal phenol-chloroform extraction
followed by ethanol precipitation of DNA was carried out as described
elsewhere \citep{Sambrook2001}.

\section{Solvent Composition and viscosity}

The solvent used in this study is the widely utilized Tris-EDTA
buffer, supplemented with NaCl. The composition of the solvent
was: 10 mM Tris, 1 mM EDTA, 0.5 M NaCl, and water.
The measured viscosity of the solvent at 20\degC\ is 1.01 mPa-s.

\section{Quantification of DNA samples}

After the DNA samples were extracted and purified, their purity were
determined using the Nano-Photometer (UV-VIS Spectrophotometer,
IMPLEN, Germany). Optical Density (O.D.) readings were taken at three
different wavelengths: 260 nm, 280 nm and 230 nm. The ratio of
absorbance at 260 nm to that of 280 nm gives a rough indication of DNA
purity \citep{Sambrook2001}. The concentrations were calculated from
absorbance reading at 260 nm (DNA shows absorption peak at 260 nm) by
Beer-Lambert's Law \citep{Sambrook2001} and also by agarose gel
electrophoresis through a serial dilution of DNA samples as suggested
elsewhere \citep{Laib20064115}. All the linear DNA samples
demonstrated $A_{260}/A_{280}$ ratio of 1.8 and above. This indicates
good purity for DNA samples, though it is largely an assumption
\citep{Laib20064115} and $A_{260}/A_{230}$ ratio from 2.0 to 2.2
(absence of organic reagents like phenol, chloroform
etc) \citep{Laib20064115}. The low molecular weight linear DNA
fragments (plasmids) were quantified through agarose gel
electrophoresis with a known standard 1 kbp DNA marker
(Fermentas). For low copy number fragments (fosmids) and very low copy
number samples (BACs), it was confirmed that the samples were not
sheared during extraction by running a very low concentration agarose
gel for extended period at low voltage. A loss of $25\%-50\%$ was
observed in the amount of DNA samples after the linearization
procedure. This is attributed to purification steps by
phenol-chloroform extraction \citep{Sambrook2001}.

\begin{figure}[tbp]  \centering
\includegraphics[width=\textwidth]{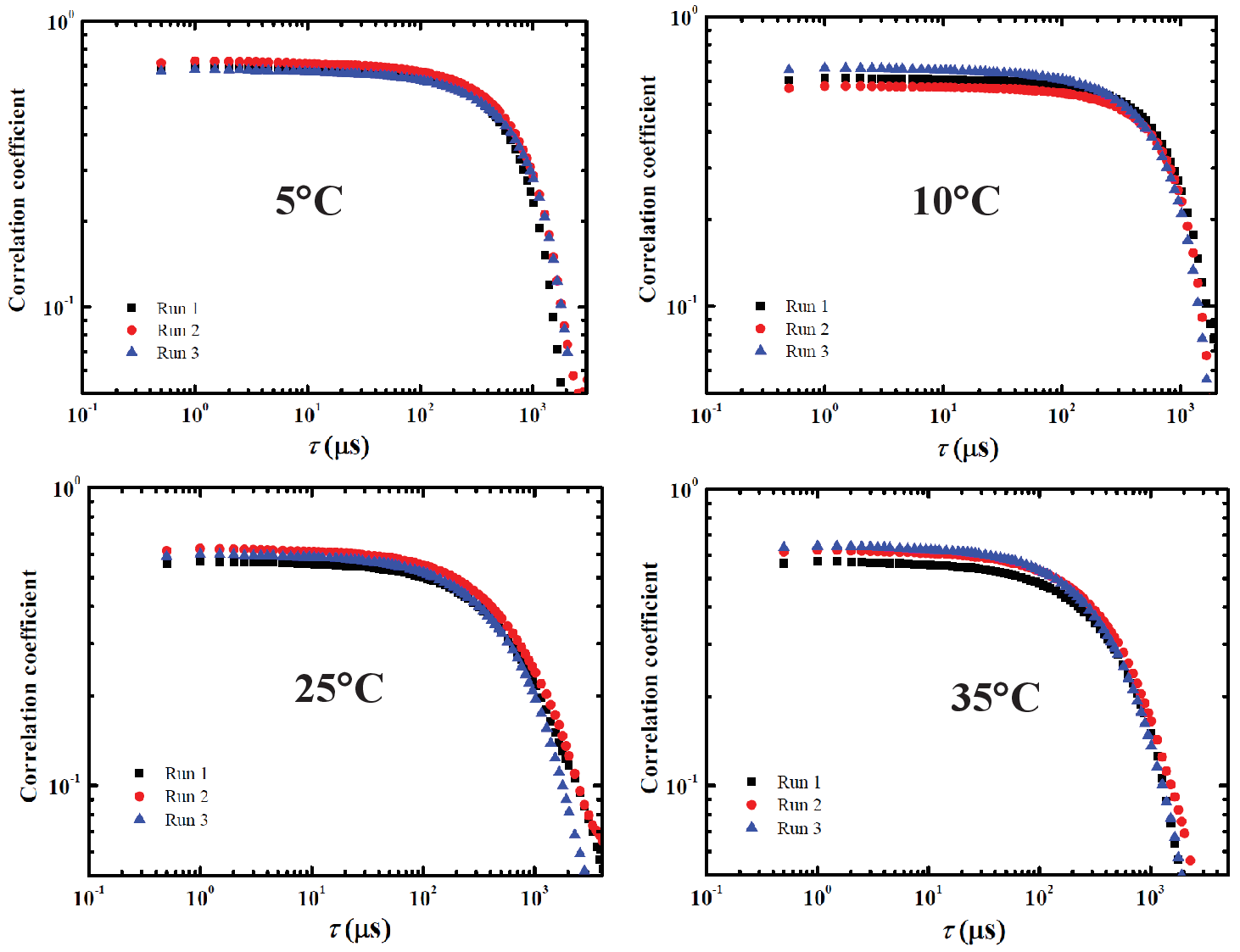}
\caption{Intensity autocorrelation spectra for 25 kbp DNA at various temperatures and $c/c^{*} = 0.1$.}
\label{fig:spectra}
\end{figure}

\section{Sample preparation for Light Scattering}

All the linear DNA fragments were dissolved individually in the
solvent and were characterized by dynamic light scattering (DLS) for
the hydrodynamic radius, \Rh\ (at the temperatures: 5\degC,
10\degC, 15\degC, 20\degC, 25\degC, 30\degC\ and 35\degC),
and by static light scattering (SLS) for the second virial coefficient (at the temperatures:
11.2\degC, 13\degC, 14\degC, 15\degC, and 20\degC)
in order to determine the $\theta$ temperature.
For both dynamic and static light
scattering, an extensive sample preparation method was followed to
ensure repeatability. The methodology of sample preparation was
modified from earlier studies \citep{Sorlie1990487, Lewis1985944,
Selis1995661} and was repeated before each measurement. The cuvette
was washed with ethanol (0.5 ml) for 5 times and kept for 15 minutes
inside laminar air flow. It was followed by wash with milliQ grade
water for $10-15$ minutes continuously. In the meantime, the solvents
were filtered with 0.45$\mu$ membrane-filter (PALL Corp.,USA) with 2
different membranes consecutively. After filtration, DNA was added to
make final concentration of $c/c^{*}$ = 0.1 (for DLS) and $c/c^{*} =
0.2$ to $0.4$ (for SLS). Note that the estimation of the overlap concentration $c^{*}$,
is discussed in section~III~A of the main paper.

\section{Dynamic Light Scattering}

\begin{figure}[tbp]  \centering
\includegraphics[width=\textwidth]{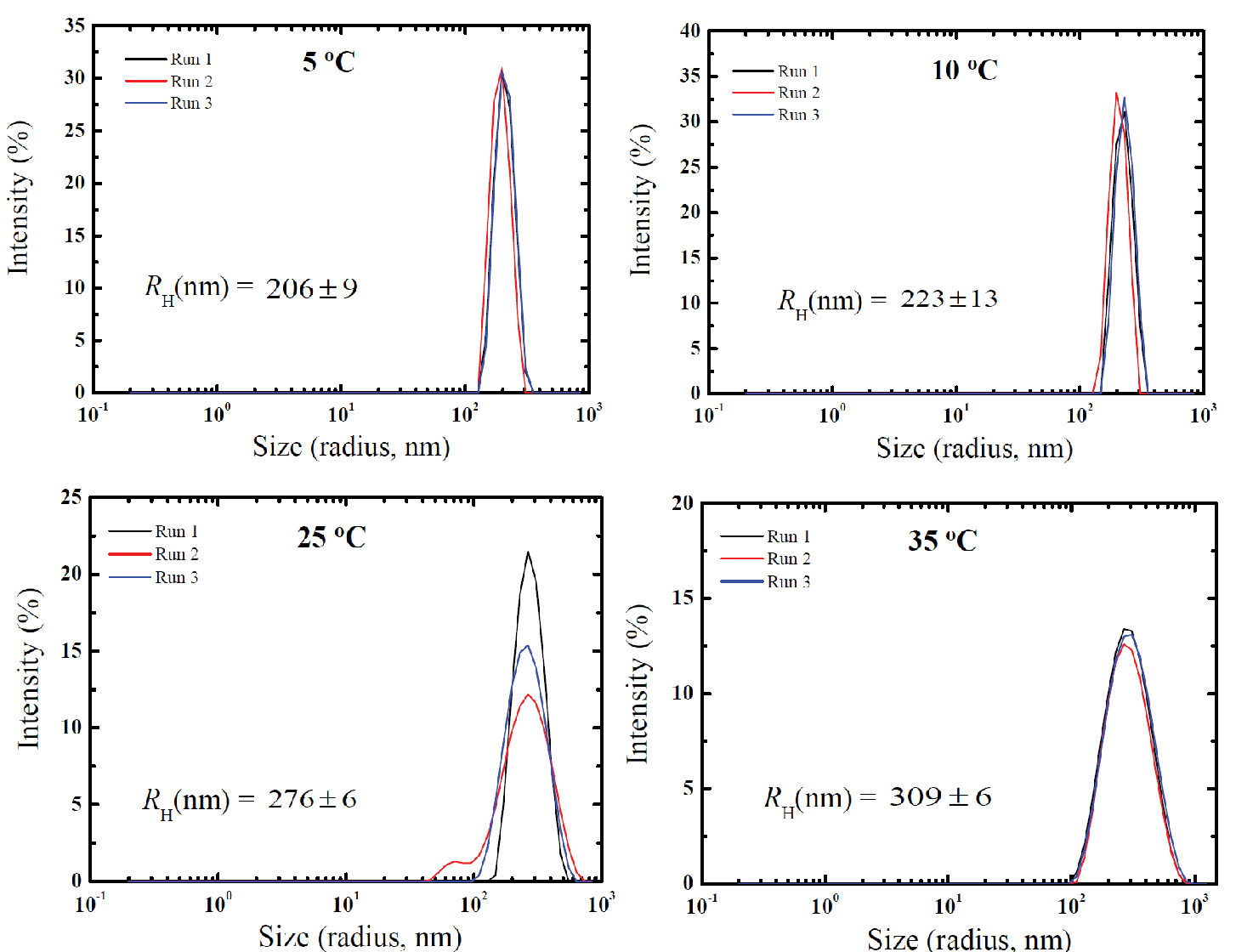}
\caption{Intensity size distributions for 25 kbp DNA at various temperatures.}
\label{fig:intdis}
\end{figure}

The hydrodynamic radius of the various DNA molecules was determined using a Zetasizer Nano ZS (ZEN3600, MALVERN, U.K.) particle size analyzer with temperature control fitted with a 633 nm He-Ne laser using back-scattering detection. This instrument uses dynamic light scattering to measure the diffusion coefficient which is then converted to an average hydrodynamic size of particles in solution using the Stokes-Einstein equation. A Standard Operating Procedure (SOP) was created using the Dispersion Technology Software (DTS 5.00, MALVERN, U.K.) to achieve the desired outcome (\Rh) without manual intervention. Scattering of the DNA solutions was measured at a fixed $173^{\circ}$ scattering angle (this enables measurements even at high sample concentrations and the effect of dust is greatly reduced). The temperature range investigated was from 5 to 35\degC. A typical example of the  ``Correlation Coefficient'', $G(\tau)= <I(t) I (t+\tau)>$ measured by the instrument for 25 kbp DNA at various temperatures and $c/c^{*} = 0.1$ is shown in \Figref{fig:spectra}. Here, $I$ is the intensity of scattered light, and $\tau$ is the time difference of the correlator. The correlation function is processed by the instrument to obtain the size distribution in terms of a plot of the relative intensity of light scattered by particles in various size classes. The Zetasizer Nano ZS has the ability to measure a wide size range (0.6--6000 nm in diameter). In this paper, we have reported sizes roughly in the range 140--2800 nm in diameter, which is within the size range of the instrument. A typical intensity distribution plot is shown in \Figref{fig:intdis} for 25 kbp DNA at 15\degC\ and 35\degC, where it can be seen that there is a single fairly smooth peak indicating the molecule's size. Readings for the size were taken in three temperature scans (a sequence of High-Low, Low-High, and High-Low temperature settings); with 5 readings at each temperature. The mean of 15 readings was taken as final hydrodynamic radius at each temperature for each DNA fragment. Measured values of \Rh\ are reported in Table~IV   of the main paper.

\begin{figure}[tbp]  \centering
\includegraphics[width=\textwidth]{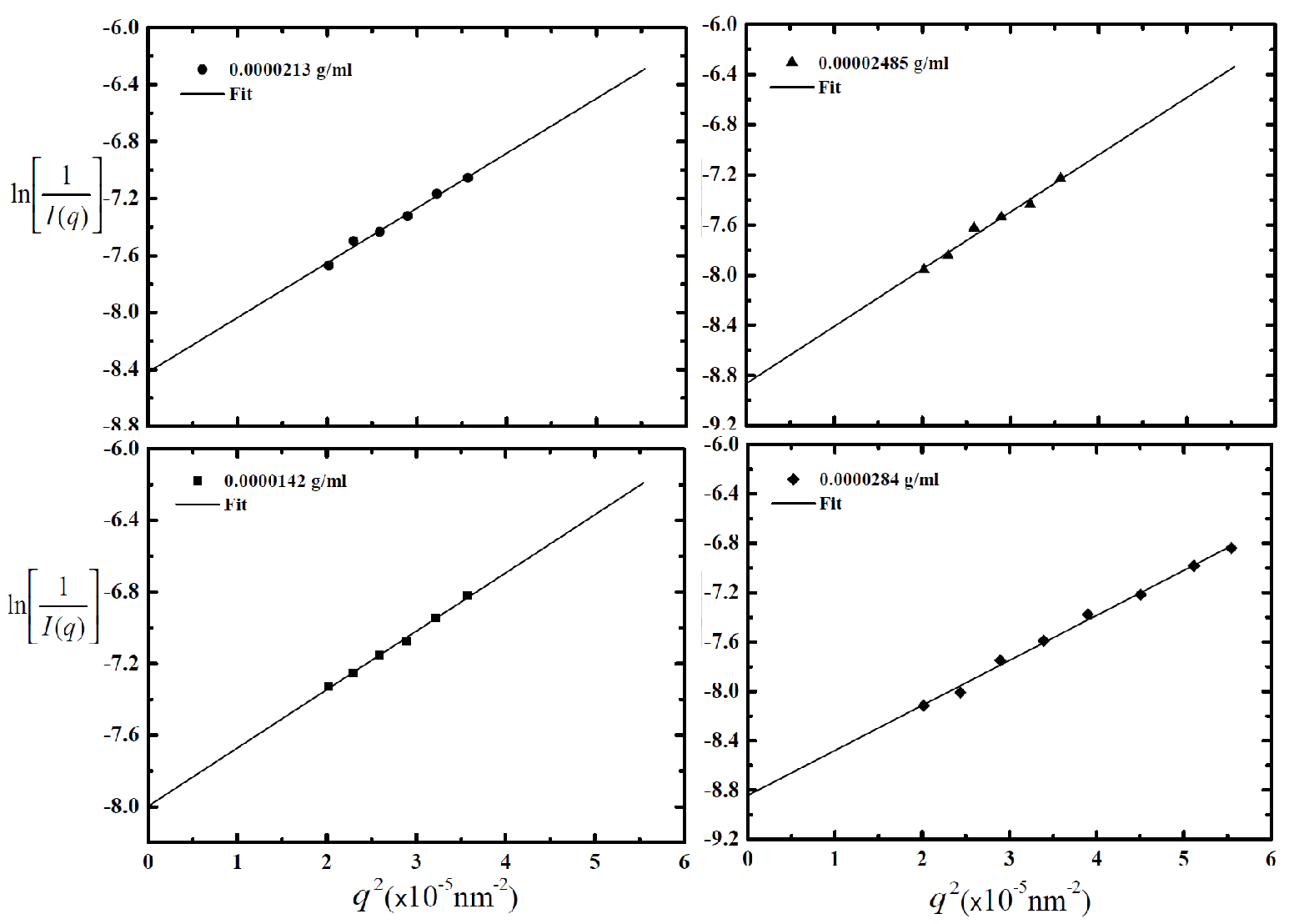}
\caption{Application of the Guinier approximation.
Intensity as a function of scattering wave vector, measured
for 25 kbp DNA at $14 \degC$ and four different concentrations,
corresponding to $c/c^{*} = 0.2, 0.3, 0.35$ and $0.4$, extrapolated to $q =0$.}
\label{fig:guinier}
\end{figure}

\section{Static Light Scattering}

The static light scattering (SLS) measurements were obtained from a
BI-200SM Goniometer (Brookhaven Instruments Corporation, USA), with a 473 nm 
wavelength Argon ion laser from Coherent, Inc. (USA), using 
BI-SLSW static light scattering software. A separate temperature
control system (PolySc, USA) was used. A single, linear,
medium molecular weight DNA fragment (25 kbp) was studied, and the
intensity of scattered light, $I (q)$ was determined as a function of the scattering vector, $q$
and polymer concentration, $c$, at 5 different temperatures between
$10$ and $20$\degC. The angle range was selected based on the sample concentration. For the highest concentration (0.0284 mg/ml) the following set of angles were used: \{15$^{\circ}$, 16.5$^{\circ}$, 18$^{\circ}$, 19.5$^{\circ}$, 21$^{\circ}$, 22.5$^{\circ}$, 24$^{\circ}$, and 25$^{\circ}$\}, while 
for the three other concentrations, the angles used were: \{15$^{\circ}$, 16$^{\circ}$, 17$^{\circ}$, 18$^{\circ}$, 19$^{\circ}$, and 20$^{\circ}$\}.
Readings were taken in two temperature scans; with 5 repeats at each
temperature. The mean of 10 repeats was taken as the final data point at
each temperature. The light scattering data was analysed according to the arguments given below in order to find the dependence of the second virial coefficient on temperature, and by this means, the $\theta$ temperature.

For solutions of macromolecules, the basic equation for the angular dependence of light scattering is the Debye-Zimm relation~\citep{Harding94, Fishman1996, RubCol03},
\begin{equation}
\frac{K c}{R_{\theta}} = \frac{1}{M} \left[ 1 + 2 \, A_{2} \, c \, M \right] \frac{1}{P(q)}
\label{debzimm}
\end{equation}
where, $K$ is an optical constant, $A_{2}$ is the second virial coefficient, $P(q)$ is the form factor, and $R_{\theta}$ is the Rayleigh excess ratio, defined by the expression,
\begin{equation}
R_{\theta} = \frac{\bar I  r^{2}}{I_{i}}
\label{rayleigh}
\end{equation}
where, $\bar I = I_\text{ex}/V$, is the excess scattered intensity $I_\text{ex}$ per unit scattering volume $V$, the quantity $I_{i}$ represents the incident intensity, and $r$ is the distance from the sample to the detector. Here, we assume that the excess scattered intensity, $I_\text{ex} = I(q) - I_{s} \approx I (q)$, since the scattered intensity $I(q)$ from the DNA solution is much greater than the scattered intensity from pure solvent, $I_{s}$. If we define the quantity,
\begin{equation}
K^{\prime} = K \left(\frac{I_{i} \, V}{r^{2}} \right)
\label{Kp}
\end{equation}
it follows from Eqs.~(\ref{debzimm}) to (\ref{Kp}) that,
\begin{equation}
\frac{K^{\prime} c}{I(q)} = \frac{1}{M} \left[ 1 + 2 \, A_{2} \, c \, M \right] \frac{1}{P(q)}
\label{debzimm2}
\end{equation}
Denoting the scattered intensity in the limit of zero scattering angle by $I_{0}$, then, since $\lim_{q \to 0} P(q) = 1$, Eq.~(\ref{debzimm2}) implies,
\begin{equation}
I_{0} \equiv \lim_{q \to 0} \, I (q) = \frac{K^{\prime} c  \, M}{\left[ 1 + 2 \, A_{2} \, c \, M \right]}
\label{I0}
\end{equation}
and, Eq.~(\ref{debzimm2}) can be rearranged in this limit to be,
\begin{equation}
\frac{c}{I_{0}}  =  \frac{1}{K^{\prime} M} + \left[\frac{2 A_{2}}{K^{\prime}} \right]  c
\label{cbyI0}
\end{equation}
If $I_{0}$ is known, then, it is clear from Eq.~(\ref{cbyI0}) that a plot of ${c}/{I_{0}}$ versus the concentration $c$ would be a straight line with intercept ${1}/({K^{\prime} M})$ and slope $({2 A_{2}}/{K^{\prime}})$. In the present instance, since we know $M$ a priori for the 25 kbp sample used in the light scattering experiments, the constant $K^{\prime}$ can be determined from the intercept. As a result, the second virial coefficient $A_{2}$ can be determined from the slope. We address the question of determining $I_{0}$ as follows.

\begin{table}[t]
\caption{The zero angle scattered intensity $I_{0}$ (in kW/cm$^{2}$) for 25 kbp DNA, at various temperatures and a range of concentrations, determined using the Guinier approximation.}
\vskip10pt
\begin{tabular}{ c  c  c  c  c  c }
\hline
$c$ ($\times$10$^{-5}$g/ml) & 11.2\degC\ & 13\degC\ & 14\degC\ & 15\degC\ & 20\degC\ \\
\hline
\hline
2.84 & 12.7$\pm$1.45 & 7.0$\pm$0.4 & 5.3$\pm$0.23 & 3.8$\pm$0.21 & 2.2$\pm$0.16 \\
\hline
2.485 & 9.3$\pm$0.72 & 5.9$\pm$0.27 & 4.4$\pm$0.1 & 3.2$\pm$0.17 & 2.1$\pm$0.1 \\
\hline
2.13 & 6.5$\pm$0.73 & 4.8$\pm$0.33 & 3.8$\pm$0.27 & 2.84$\pm$0.089  & 1.54$\pm$0.089  \\
\hline
1.42 & 3.2$\pm$0.29 & 2.8$\pm$0.33 & 2.4$\pm$0.15 & 1.9$\pm$0.15 & 1.08$\pm$0.058 \\
\hline
\end{tabular}
\label{tab:I0ds1}
% \vskip-10pt
\end{table}

From Eqs.~(\ref{debzimm2}) and~(\ref{I0}), it follows that,
\begin{equation}
\frac{ I_{0}}{I (q)} = \frac{1}{P(q)}
\label{onebyP}
\end{equation}
At low scattering angles, $q^{2}R_{g}^{2} \sim \mathcal{O} (1)$, the form factor is often approximated by the Guinier function~\citep{RubCol03},
\begin{equation}
P(q) = \exp \left[  - \frac{q^{2} R_{g}^{2}}{3}  \right]
\label{gui}
\end{equation}
It then follows from Eq.~(\ref{onebyP}) that,
\begin{equation}
\ln \left( \frac{1}{I (q)} \right) = \ln \left( \frac{1}{I_{0}} \right) +   \left( \frac{R_{g}^{2}}{3} \right) \, q^{2}
\end{equation}
As a result, a plot of $\ln ({1}/{I (q)})$ versus $q^{2}$ would be linear, and the zero angle scattered intensity $I_{0}$ could be determined from the intercept without a knowledge of either $K^{\prime}$, $A_{2}$ or $R_{g}$.

Figure~\ref{fig:guinier} displays the intensity as a function of scattering wave vector for 25 kbp DNA, at $14 \degC$ and four different concentrations (corresponding to $c/c^{*} = 0.2, 0.3, 0.35$ and $0.4$), plotted semilog. The fact that nearly all the measured intensity data, for the various values of $q^{2}$, lies on the fitted lines indicates that the Guinier is a good approximation in this case. All the values of $I_{0}$, determined by extrapolating linear fits of $\ln ({1}/{I (q)})$ versus $q^{2}$ data to $q=0$, at various temperatures and concentrations, are listed in Table~\ref{tab:I0ds1}.

Attempts to use an alternative procedure to find $I_{0}$ by assuming that $P(q)$ is a linear function of $q^{2}$, for $q^{2}R_{g}^{2} \lesssim \mathcal{O} (1)$, i.e., $P(q) = 1- (q^{2}R_{g}^{2}/3)$ and plotting $1/I (q)$ versus $q^{2}$ (as in a Zimm plot), or $\sqrt{1/I (q)}$ versus  $q^{2}$ (as in a Berry plot)~\citep{Burchard2008} and extrapolating the fitted line through the data to $q = 0$, did not lead to consistent results in the subsequent analysis.

\begin{figure}[tbp]  \centering
\includegraphics[width=0.7\textwidth]{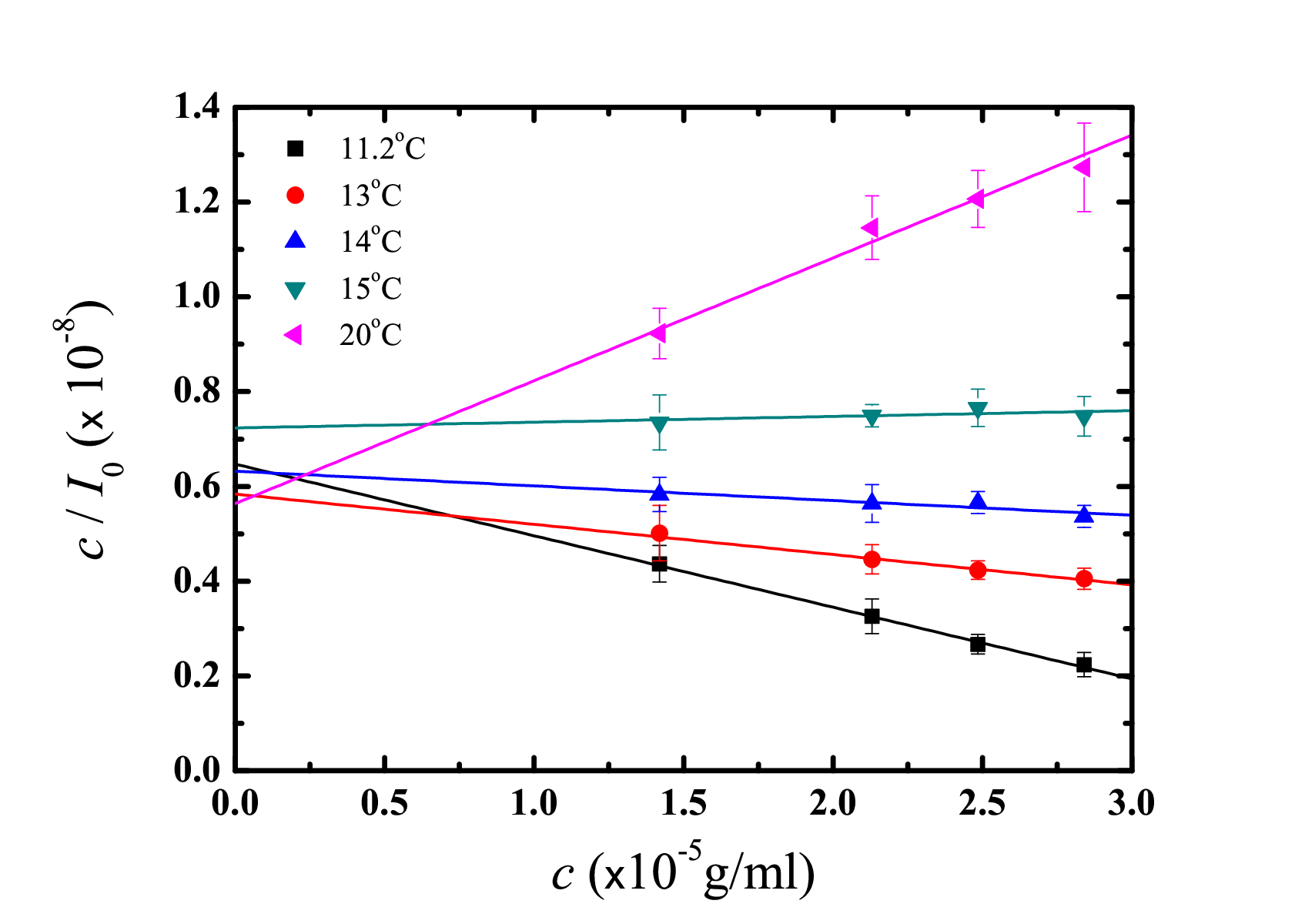}
\caption{Linear dependence of the ratio $c/I_{0}$ on concentration, as expressed by Eq.~(\ref{cbyI0}), for 25 kbp DNA at various temperatures. The constant $K^{\prime}$ is determined from the intercept, and the temperature dependence of the second virial coefficient is determined from the slope. }
\label{fig:A2andK}
\end{figure}

Once $I_{0}$ is known, the ratio $c/I_{0}$ can be plotted versus $c$, and both $K^{\prime}$ and $A_{2}$ determined, as discussed below Eq.~(\ref{cbyI0}). Figure~\ref{fig:A2andK} is a plot of $c/I_{0}$ versus $c$ for the values of $I_{0}$ listed in Table~\ref{tab:I0ds1}, at the various temperatures at which measurements were carried out. While in principle the data for all the temperatures should extrapolate to a unique intercept at $c=0$, the scatter observed in Fig.~\ref{fig:A2andK} reflects the uncertainty in the $I(q)$ data. Accounting for the spread in the values of the intercepts, leads to the value, $K^{\prime} = 9.5 \pm 0.3$ (mol~cm~watts/g$^{2}$). The dependence on temperature of the second virial coefficient $A_{2}$, determined from the slopes of the fitted lines in Fig.~\ref{fig:A2andK}, and the subsequent analysis, is discussed in Appendix~A of the main paper.

\section{Determination of the chemistry dependent constant $k$}

\begin{figure}[tbp]
\centering
\includegraphics[width=0.7\textwidth]{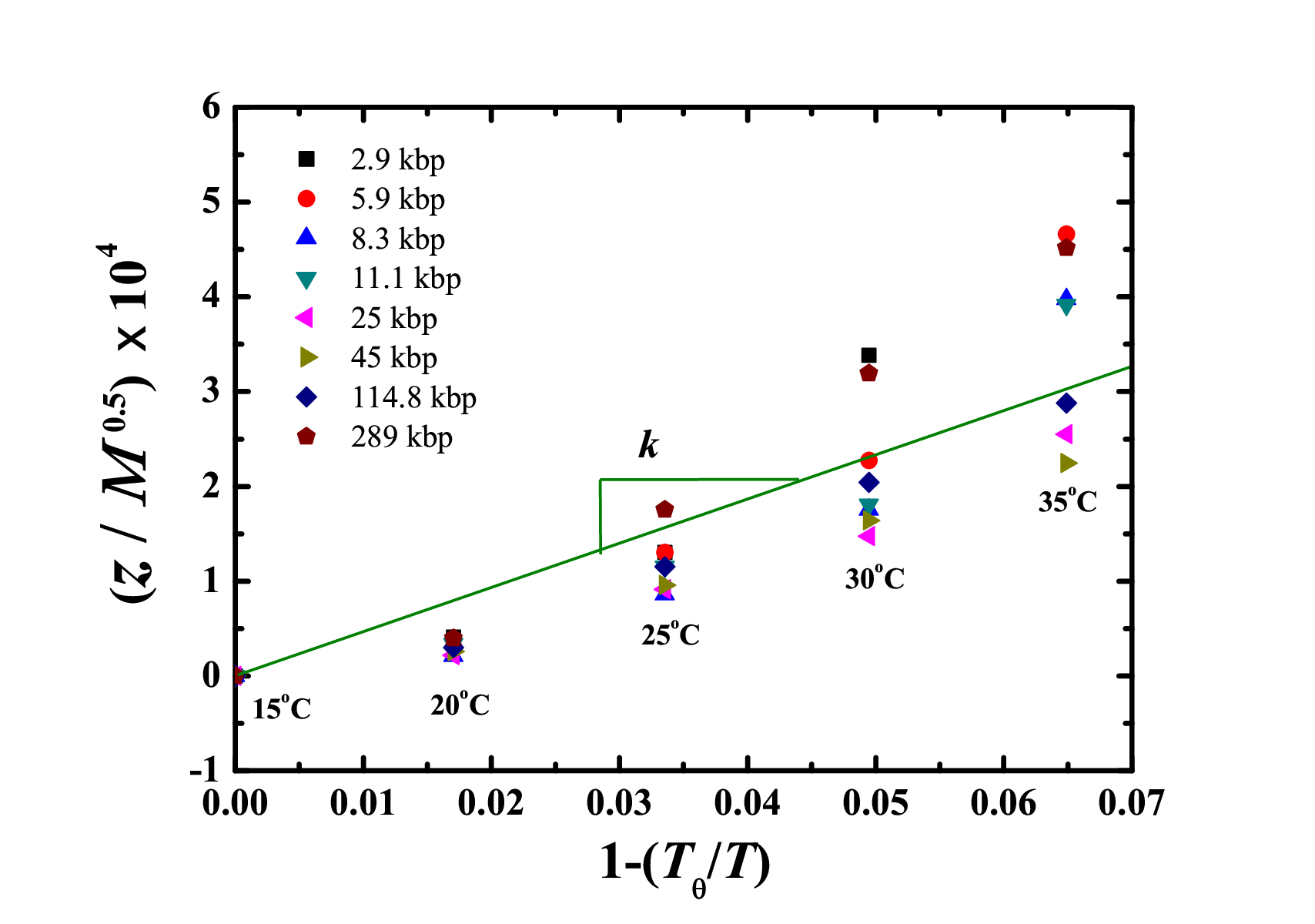}
%\gpfig{k.eps}
  \caption[Determination of k]{\label{kfig} Determination of the chemistry dependent constant $k$.
  The data points are least square fitted with a straight line and the slope of this line gives
    $k$ \citep{Kumar20037842}.}
\end{figure}

The value of the chemistry dependent constant $k$ (appearing in the definition of the solvent quality parameter $z$)  has been determined for the solvent by adopting a procedure elaborated in an earlier
work~\citep{Kumar20037842}, and briefly summarized in the main text (see Appendix~B).
\Figref{kfig} displays plots of $f^{-1}_\text{H}(\ah^\text{expt})/\sqrt{M}$ versus $\hat \tau$ for the
solvent used in this study. Only the temperatures above the theta point are considered here.
The data points were least square fitted with a straight line, and the slope $k$ determined.
A value of 15\degC\ has been used for the $\theta$-temperature. The value of $k$ found by this procedure is
$0.0047 \pm 0.0003$ (g/mol)$^{-1/2}$.

\bibliography{jorrefs}
\noindent
\bibliographystyle{JORnat}